\documentclass[11pt]{article}
\pdfoutput=1
\usepackage{amsmath,amssymb,esint,graphicx,subfig,float,afterpage,lipsum,amsthm,amsfonts,dsfont,color,tikz}
\usepackage[toc,page]{appendix}
\usepackage{mathrsfs}
\usepackage{tensor}
\usepackage{braket}
\usepackage{cite}
\usepackage{caption}
\usepackage{url}

\usepackage[a4paper,width=145mm,top=30mm,bottom=30mm]{geometry}
\numberwithin{equation}{section}

\usepackage[colorlinks=true, linkcolor=blue, bookmarks=true]{hyperref}

\usepackage{subfig}

\newcommand{\extd}{d}
\newcommand{\st}{t}

\DeclareMathOperator\arcsinh{arcsinh}

\renewcommand{\title}[1]{\vbox{\center\bf{\Large{#1}}}\vspace{5mm}}
  \renewcommand{\author}[1]{\vbox{\center#1}\vspace{5mm}}
  \newcommand{\address}[1]{\vbox{\center\em#1}}
  \newcommand{\email}[1]{\vbox{\center\tt#1}\vspace{5mm}}

\begin{document}
\begin{titlepage}
\begin{center}
\hfill \\
\hfill \\
\vskip 1cm
\title{Slow scrambling in extremal BTZ and microstate geometries}
\author {Ben~Craps,$^{1}$ Marine~De~Clerck,$^{1}$ Philip~Hacker,$^{1}$ K\'evin~Nguyen,$^{2}$
Charles~Rabideau$^{1}$}
\address{\vspace{2mm}
$^{1}$Theoretische Natuurkunde, Vrije Universiteit Brussel (VUB) and\\
The International Solvay Institutes, Pleinlaan 2, B-1050 Brussels, Belgium \vspace{1mm}\\
$^{2}$Black Hole Initiative, Harvard University,\\ Cambridge, MA 02138, USA \\}
\end{center}
\vspace{5mm}
\begin{abstract}
Out-of-time-order correlators (OTOCs) that capture maximally chaotic properties of a black hole are determined by scattering processes near the horizon. This prompts the question to what extent OTOCs display chaotic behaviour in horizonless microstate geometries. This question is complicated by the fact that Lyapunov growth of OTOCs requires nonzero temperature, whereas constructions of microstate geometries have been mostly restricted to extremal black holes.

In this paper, we compute OTOCs for a class of extremal black holes, namely maximally rotating BTZ black holes, and show that on average they display ``slow scrambling'', characterized by cubic (rather than exponential) growth. Superposed on this average power-law growth is a sawtooth pattern, whose steep parts correspond to brief periods of Lyapunov growth associated to the nonzero temperature of the right-moving degrees of freedom in a dual conformal field theory.

Next we study the extent to which these OTOCs are modified in certain ``superstrata'', horizonless microstate geometries corresponding to these black holes. Rather than an infinite throat ending on a horizon, these geometries have a very deep but finite throat ending in a cap. We find that the superstrata display the same slow scrambling as maximally rotating BTZ black holes, except that for large enough time intervals the growth of the OTOC is cut off by effects related to the cap region, some of which we evaluate explicitly.
\end{abstract}

\vfill
\email{Ben.Craps, Marine.Alexandra.De.Clerck, Philip.Hacker, Charles.Rabideau@vub.be and kevin\_nguyen@g.harvard.edu}
\end{titlepage}

\tableofcontents

\section{Introduction}
The nature of the microstates of black holes in quantum gravity is a matter of ongoing debate. Some take recent progress on recovering the Page curve of evaporating black holes from semiclassical gravity \cite{Penington:2019npb,Almheiri:2019psf,Penington:2019kki,Almheiri:2019qdq} (see \cite{Almheiri:2020cfm} for a review) as evidence that black hole microstates have a smooth horizon and an interior region. Others are convinced that many or all black hole microstates correspond to horizonless geometries (see \cite{Mathur:2009hf} for a review), and point to explicit constructions of increasingly large families of such geometries in string theory (see e.g.\ \cite{Bena:2016ypk}).

If black hole microstates correspond to horizonless geometries, one expects that observables computed in such geometries should approximately reproduce those computed in the ``na\"ive’’ black hole geometry. Nevertheless, sufficiently accurate computations or measurements should be able to distinguish them. Examples include studies of the approximate thermality of probes in ensembles of gravitational microstates \cite{Balasubramanian:2005mg,Balasubramanian:2007qv}, an analogue of Hawking radiation in special microstate geometries of a non-extremal black hole \cite{Chowdhury:2007jx}, and the behaviour of retarded two-point functions in certain microstates of extremal black holes \cite{Bena:2019azk}.

Black hole horizons have played a central role in recent connections between gravity and quantum chaos. Out-of-time-order correlators (OTOCs) in holographic field theories in a thermal ensemble display transient Lyapunov growth, with Lyapunov exponent equal to $2\pi$ times the temperature \cite{Shenker:2013pqa}. Such systems have been argued to be maximally chaotic \cite{Maldacena:2015waa}. In the dual gravitational description, the OTOC corresponds to a scattering process very close to a black hole horizon. In contrast to quasinormal mode (QNM) decay, which depends on what happens within a few Schwarzschild radii of the horizon, chaos probes what happens {\em at} the horizon \cite{Polchinski:2015cea}. This makes it very interesting to investigate how OTOCs distinguish between black hole geometries and horizonless microstate geometries.

An important complication is that so far constructions of microstate geometries that closely resemble black holes have been mostly restricted to extremal black holes, which are under better control due to supersymmetry. Such black holes have zero temperature, which requires modification of the above discussion. While it would be interesting from various points of view to have microstate geometries corresponding to non-extremal black holes, constructing them is not an easy task. In the present paper, we will therefore extend computations of chaos as measured by OTOCs to the case of certain extremal black holes and associated microstate geometries.

To set the stage, we briefly review some aspects of quantum chaos and OTOCs, mainly following \cite{Polchinski:2015cea}. In classically chaotic systems, neighboring phase space trajectories diverge exponentially, $\partial q(t)/\partial q(0)\sim \exp(\lambda t)$. Since $\partial q(t)/\partial q(0)=\{q(t),p(0)\}$, this motivates the study in quantum mechanics of commutators of operators at different times. More specifically, one is interested in $-\langle [V(0),W(t)]^2\rangle_\beta$, which we will refer to as the commutator squared, and where $\beta$ is the inverse temperature.
\begin{figure}[h]
\includegraphics[width=9cm]{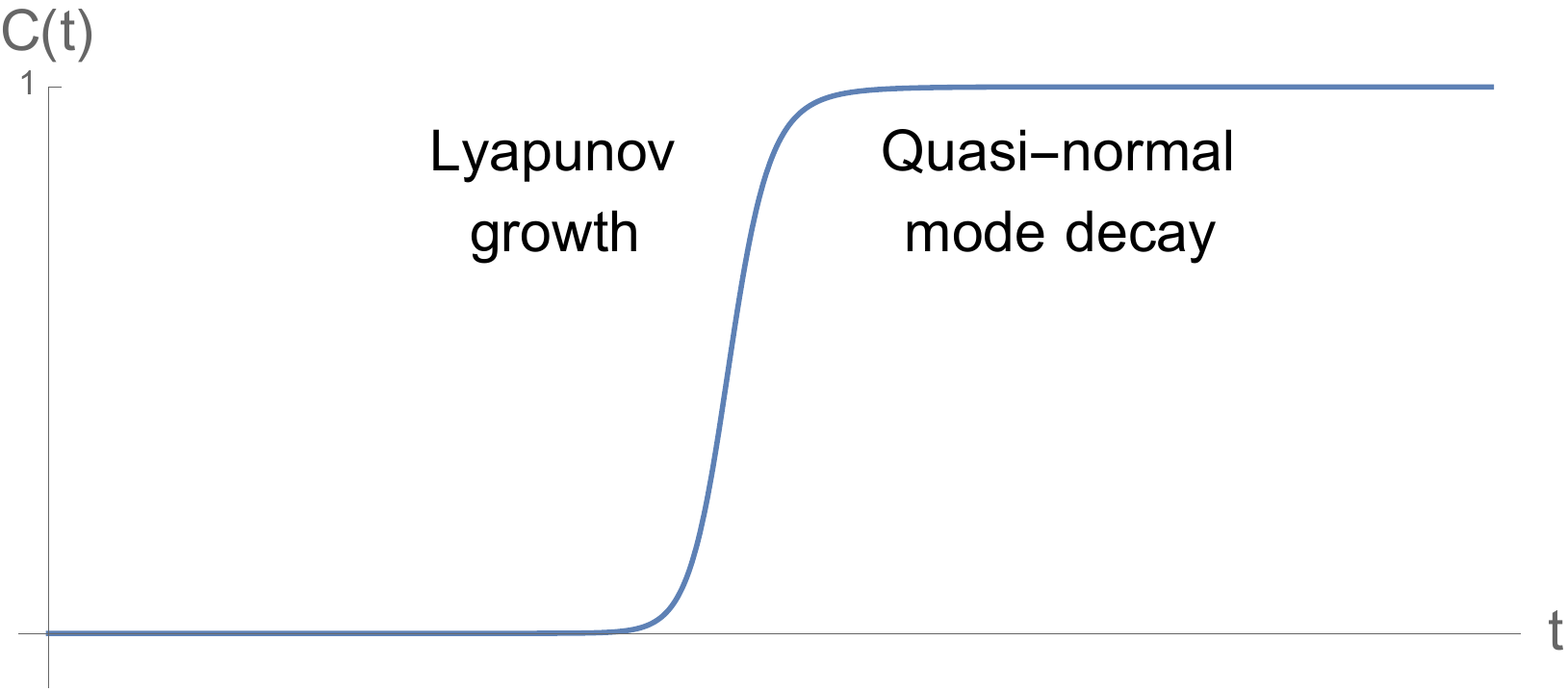}
\centering
\caption{Schematic plot (based on \cite{Polchinski:2015cea}) of the expected behaviour of the normalized commutator squared $C(t)$ over time.}
\label{fig:OTOC1}
\end{figure}
As displayed in figure~\ref{fig:OTOC1}, in theories with a chaotic semi-classical limit and for operators that commute at $t=0$ this quantity typically displays two exponential behaviours, namely transient Lyapunov growth followed by saturation, where the latter is described by Ruelle resonances. Holographically, Ruelle resonances correspond to quasinormal mode decay, also visible in two-point functions, while the Lyapunov growth is due to near-horizon blueshifts, which manifest themselves in 2-to-2 scattering amplitudes that can be associated to OTOCs \cite{Shenker:2014cwa}.
For times large compared to the inverse temperature, the normalized commutator squared $C(t)$
is simply given by 1 minus the real part of the out-of-time-order correlator,
\begin{align}
C(t) &\equiv \frac{-\langle [V(0),W(t)]^2\rangle_\beta}{2\langle VV\rangle_\beta\langle WW\rangle_\beta}\approx 1- {\rm Re} \, {\rm OTOC}(t)  \ \ \ \  (t\gg\beta)\,,
\end{align}
where the normalized OTOC is defined as
\begin{align}
{\rm OTOC}(t) &\equiv \frac{\langle V(0)W(t)V(0)W(t)\rangle_\beta}{\langle VV\rangle_\beta\langle WW\rangle_\beta} \,.
\end{align}
This is due to the fact that contributions like $\langle V(0)W(t)W(t)V(0)\rangle_\beta$ can be interpreted as the 2-point function of $W(t)$ in a state created by acting with $V(0)$ on the thermal state; this state behaves thermally after one waits a few thermal times, so this contribution factorizes.

In fact, the commutator squared contains more structure than shown in figure~\ref{fig:OTOC1}. In a 2d holographic Conformal Field Theory (CFT), if the conformal dimensions satisfy
$h_W \gg h_V \gg 1$ and the time is large compared to both the spatial separation and the inverse temperature, the OTOC was computed in \cite{Roberts:2014ifa},
\begin{equation}
\label{eqn:Stanford_Roberts_OTOC}
\frac{\langle V(i\epsilon_1,x)W(t+i\epsilon_3,0)V(i\epsilon_2,x)W(t+i\epsilon_4,0)\rangle_\beta}{\langle V(i\epsilon_1,0)V(i\epsilon_2,0)\rangle_\beta\langle W(i\epsilon_3,0)W(i\epsilon_4,0)\rangle_\beta}\approx \left(
\frac{1}{1-\frac{24\pi ih_W}{\epsilon_{12}^*\epsilon_{34}c}e^{\frac{2\pi}{\beta}(t-|x|)}}
\right)^{2h_V}.
\end{equation}
Here $c$ is the central charge, which is large, and
\begin{equation}
    \epsilon_{ij}=i\left(e^{\frac{2\pi}{\beta}i\epsilon_i}-e^{\frac{2\pi}{\beta}i\epsilon_j}
    \right).
\end{equation}
At sufficiently early times, one gets Lyapunov behaviour from the $1/c$ expansion,
\begin{align}
\frac{\langle V(i\epsilon_1,x)W(t+i\epsilon_3,0)V(i\epsilon_2,x)W(t+i\epsilon_4,0)\rangle_\beta}{\langle V(i\epsilon_1,0)V(i\epsilon_2,0)\rangle_\beta\langle W(i\epsilon_3,0)W(i\epsilon_4,0)\rangle_\beta}
\approx 1 +  \frac{48 \pi i h_W h_V}{\epsilon^*_{12} \epsilon_{34} c} e^{\frac{2\pi}{\beta}{(t-|x|)}} \,.
\end{align}
Note that this exponential growth is suppressed by a prefactor that is small in the large-$c$ semi-classical limit. This growth will persist until it competes with the small prefactor at the scrambling time
\begin{align}
t_s = |x| + \frac{\beta}{2\pi} \log \frac{|\epsilon^*_{12} \epsilon_{34}| c}{48 \pi  h_W h_V}\,,
\end{align}
which is the time at which the commutator squared, $C(t)$, first becomes $O(1)$.

\begin{figure}[h]
\includegraphics[width=12cm]{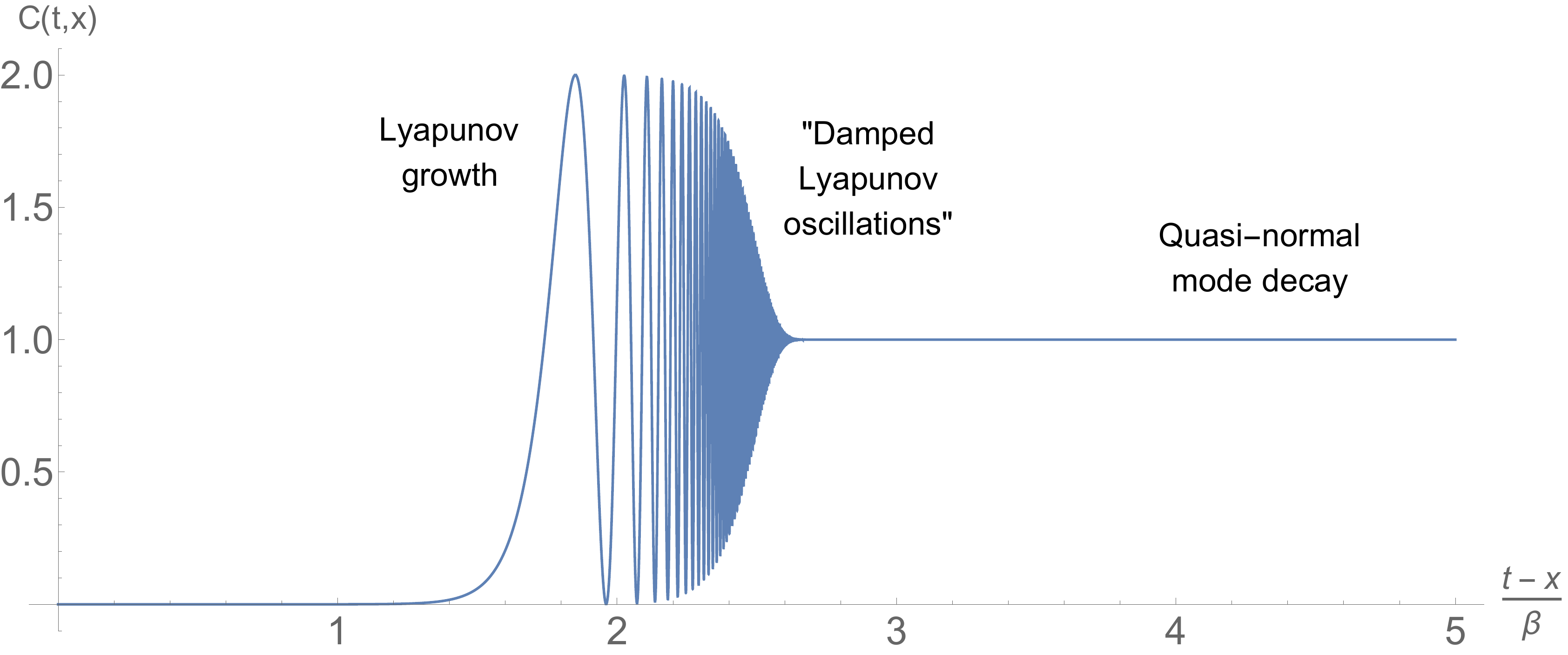}
\centering
\caption{Behaviour of the normalized commutator squared in a 2d holographic CFT. In this plot, $\frac{h_W}{c \epsilon^*_{12}\epsilon_{34}}=\exp(-24.5)$ and $h_V=\exp(9)$.}
\label{fig:OTOC2}
\end{figure}

After the scrambling time there is a region of oscillatory behaviour, which subsequently decays away. This decay is also controlled by the Lyapunov exponent and occurs well before the quasi-normal regime.
Nonetheless, at sufficiently late times one sees the faster exponential quasi-normal mode decay in the tail of this decay. The details of this intermediate regime are described in more detail in appendix \ref{sec:OTOC_oscillations} and the behaviour of the OTOC described by \eqref{eqn:Stanford_Roberts_OTOC} is depicted in figure~\ref{fig:OTOC2}.

At very late times, the quasi-normal mode decay is expected to stop as the OTOC cannot continue to decay forever in a  unitary theory \cite{Maldacena:2001kr}. The result \eqref{eqn:Stanford_Roberts_OTOC} was derived assuming that only the contribution from the Virasoro conformal block of the identity is important and this approximation is expected to breakdown at very late times \cite{Fitzpatrick:2015dlt,Fitzpatrick:2016ive,Fitzpatrick:2016mjq,Bombini:2017sge}. 

To get a feeling for what to expect for extremal black holes, which involve zero temperature, we first review what happens in vacuum. In \cite{Roberts:2014ifa} it was found that for infinite $\beta$ and $t\gg |x|$
\begin{equation}
\label{eq:Stanford Roberts 2}
\frac{\langle V(i\epsilon_1,x)W(t+i\epsilon_3,0)V(i\epsilon_2,x)W(t+i\epsilon_4,0)\rangle}{\langle V(i\epsilon_1,0)V(i\epsilon_2,0)\rangle \langle W(i\epsilon_3,0)W(i\epsilon_4,0)\rangle}
 \approx \left(
\frac{1}{1-\frac{24\pi ih_W}{c(\epsilon_1-\epsilon_2)(\epsilon_3-\epsilon_4)}(t-|x|)^2}
\right)^{2h_V}\,.
\end{equation}
At early times, we again see a period of growth suppressed by a small prefactor,
\begin{equation}
\left(
\frac{1}{1-\frac{24\pi ih_W}{c(\epsilon_1-\epsilon_2)(\epsilon_3-\epsilon_4)}(t-|x|)^2}
\right)^{2h_V}
\approx 1+\frac{48\pi i h_Vh_W}{c(\epsilon_1-\epsilon_2)(\epsilon_3-\epsilon_4)} (t-|x|)^2 \,.
\end{equation}
This result for the OTOC is qualitatively similar to the finite temperature case, except that the exponential Lyapunov growth is replaced by quadratic growth in time. The link between the OTOC and the commutator squared is now less straightforward, because contributions like $\langle V(0)W(t)W(t)V(0)\rangle$ need not factorize at zero temperature.

In this paper, we will focus on maximally rotating BTZ black holes and some of their microstate geometries. In the dual CFT, the left-movers are at zero temperature while the right-movers are at finite temperature $T_R$. 
To guide our expectations for computing OTOCs in BTZ geometries, we refer to earlier studies of non-maximally rotating BTZ black holes, including \cite{Jahnke:2019gxr, Mezei:2019dfv}, where Lyapunov growth alternates between the left and right-moving temperatures \cite{Mezei:2019dfv} at small time scales, yet the overall growth is controlled by the Bekenstein--Hawking temperature. These results cannot be applied directly to the extremal case, since they assume the regime where $t \gg \beta$ whereas $\beta$ diverges in the extremal limit. Yet our results are compatible with extrapolating their conclusion to our setting, since for sufficiently high right-moving temperature one finds a small sawtooth-like modulation on top of power law growth, where the steep parts of the sawtooth correspond to brief periods of Lyapunov growth at the nonzero right-moving temperature. Nonetheless, on average the scrambling is slow. It is worth emphasizing that the sawtooth-like modulation is tied to the compactness of the spatial direction of the dual CFT. In contrast, the results \eqref{eqn:Stanford_Roberts_OTOC} and \eqref{eq:Stanford Roberts 2} obtained in \cite{Roberts:2014ifa} were derived in the decompactified limit such that they do not display such a sawtooth-like modulation.

In \cite{Poojary:2018esz}, the gravitational modes responsible for scrambling corresponding to the zero left moving temperature and the non-zero right moving temperature were identified using an effective action. However, as emphasised in \cite{Mezei:2019dfv}, identifying the overall rate of growth relevant for scrambling requires the detailed computation of the OTOC we present in this work.
The instantaneous Lyapunov exponent in rotating ensembles has been bounded in \cite{Halder:2019ric}.
Scrambling in BTZ has also been studied from the perspective of mutual information in \cite{Stikonas:2018ane}.
In the context of microstate geometries resembling maximally rotating BTZ black holes, note that an interesting recent paper \cite{Bianchi:2020des} has described a different kind of Lyapunov behaviour associated to geodesic instability near photon spheres. This latter Lyapunov exponent is related to quasi-normal decay \cite{Cardoso:2008bp}, while the focus of our work is on the Lyapunov growth displayed by OTOCs, which is of a different nature.

A method to compute OTOCs within AdS/CFT, which is based on the geodesic approximation to the propagation of bulk fields in asymptotically AdS spacetimes, has been developed in \cite{Balasubramanian:2019stt}. We give a brief summary thereof in section~\ref{section:geodesic}.
In practice, this method requires one to consider a particle falling in from the earlier boundary insertion point of the OTOC, together with the linearised gravitational shock wave that it sources.
Similarly, an outgoing particle reaching the later boundary insertion point needs to be considered. Within the geodesic approximation, the OTOC is then determined by
\begin{equation}
\label{eq:OTOC-delta}
\text{OTOC}\sim e^{i\delta},
\end{equation}
where the \textit{eikonal phase} $\delta$ encodes the interaction of each particle with the gravitational shock wave emitted by the other one.
Compared to the original method put forward by Shenker and Stanford \cite{Shenker:2014cwa} or follow up works such as \cite{Jahnke:2019gxr,Mezei:2019dfv}, this approach allows us to work at zero temperature. The difference is that the original method makes the approximation, well motivated at finite temperature, that the shock wave propagates exactly on the horizon. However, this approximation clearly does not apply in the vacuum where there is no horizon. The applicability of the new approach summarized in section~\ref{section:geodesic} to the zero temperature case was demonstrated in \cite{Balasubramanian:2019stt} where the quadratic growth associated to \textit{slow scrambling} in vacuum was obtained, finding agreement with earlier CFT results \cite{Roberts:2014ifa}. Similarly, the results of \cite{Mezei:2019dfv} show that the shock wave diverges in the extremal limit when the approximation of placing it on the horizon is made.
Therefore, working with a method valid at zero temperature which accurately computes the shock wave without making the approximation of placing it on the horizon, is again crucial to the study of extremal BTZ and microstate geometries which we initiate in this paper.

The gravitational scattering amplitude of highly energetic particles turns out to be generically proportional to the corresponding center-of-mass energy. Hence, at finite temperature a rough estimate of the time-dependence of an OTOC may be obtained by computing this simple quantity; see also \cite{videomaldacena}. We will find that the story is more subtle in the zero temperature case, nevertheless the center-of-mass energy gives useful intuition in instances where gravitational shock waves are difficult to compute, as is the case when the source particles propagate in microstate geometries. In appendix \ref{app:center-of-mass}, we compute the center of mass energy in rotating BTZ and find an early time period of power-law growth. This early time period persists for longer and longer as we approach the extremal BTZ limit, leading to slow scrambling.

\begin{figure}[t]    \centering
    \includegraphics[width=8cm]{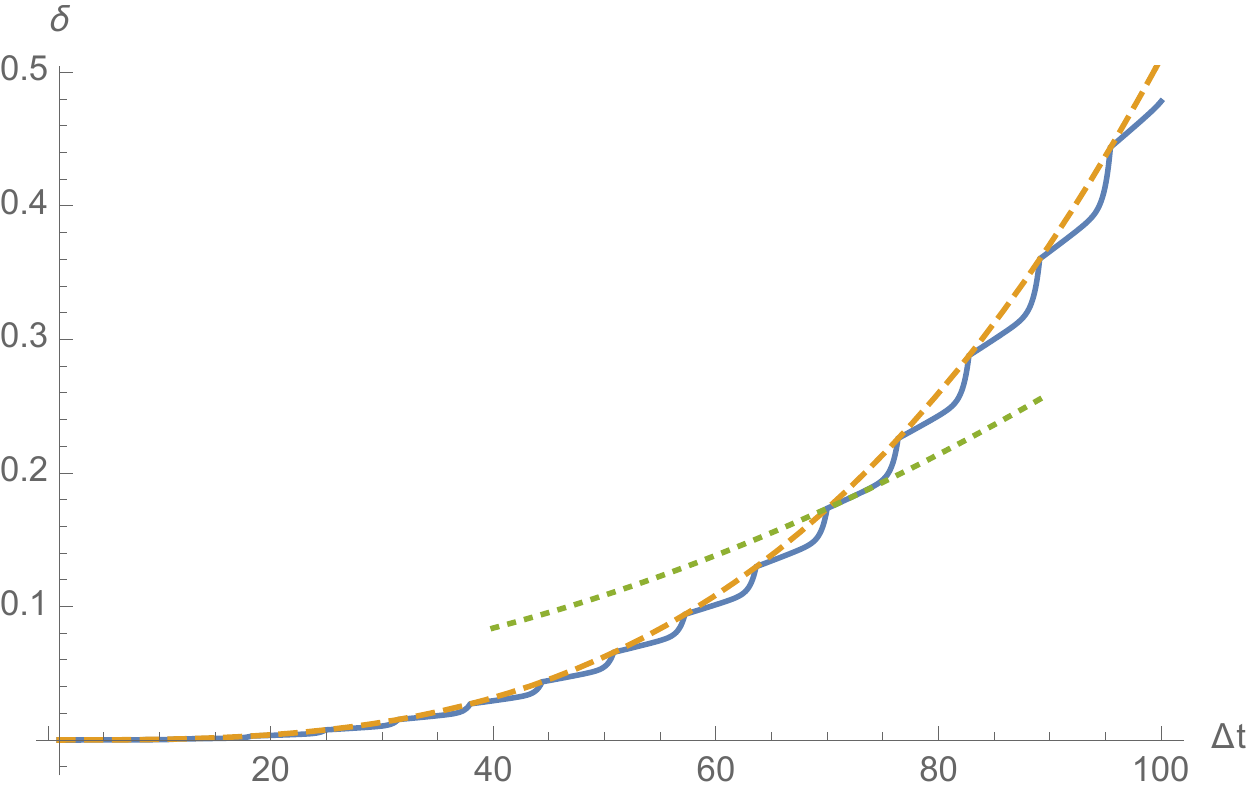}    \caption{The growth of the eikonal phase $\delta$, related to the OTOC through \eqref{eq:OTOC-delta}, of two scalar operators at a fixed spatial separation in extremal BTZ, computed within the geodesic approximation (solid blue line). The growth is cubic on average and can be very well approximated by the functional dependence depicted in orange (dashed), which scales as $\sim \frac{4 r_+^2 t^3}{3 \pi }-\frac{2 r_+}{\pi}t^2 \log \left( 4r_+ t \right)+2 r_+^2 t^2$. On top of this average growth, we find a sawtooth-like modulation that starts with a quadratically growing piece, followed by an exponential growth that catches up with the overall $t
   ^3$ growth. One of the phases of $t
   ^2$ growth is depicted by superimposing the dashed green (dotted) line, obtained by summing the quadratic contributions of a fixed 20 images. This figure is discussed in detail in section~\ref{subsection: eikonal phase}.}
    \label{fig:OTOC_cubic}\end{figure}

In section~\ref{section:BTZ} we provide a detailed holographic computation of the OTOC of scalar  operators within the geodesic approximation, in CFT states dual to maximally rotating BTZ black holes. First, we review the properties of geodesics in this spacetime. Then we consider the shock waves emitted by the scattered particles along their trajectories,  computed in appendix~\ref{appendix:shock wave}.
We deal with the angular periodicity in BTZ by first computing the shock wave in the black brane geometry and then using the method images to find the periodic solution. We find that a large but finite number of these images contribute to the shock wave at any given time and that the number of images grows linearly with time. This explains the divergence found when taking the zero-temperature limit of the result found using the earlier approach -- the late time approximation necessary to place the shock wave on the horizon leads to a diverging sum over images.
This sum over images yields an enhancement factor in the strength of the gravitational interaction which, controlled by the center-of-mass energy of the interaction, would otherwise have grown quadratically with time.
All in all, we find that the OTOC displays approximately cubic growth in time, with a sawtooth-like modulation that alternates between quadratic and exponential growth. In addition, the latter exponential Lyapunov growth is associated to the nonzero right-moving temperature $T_R$. This is displayed in figure~\ref{fig:OTOC_cubic}. The scrambling associated to scalar operators is therefore slow on average. In particular, for the scrambling time defined as the time at which the eikonal phase $\delta$ becomes $O(1)$, we obtain
\begin{align}
\label{eq:ts-intro}
      t_s &\simeq\left( \frac{c\, \varepsilon^2}{16 h_V h_W } \right)^{\frac13}\,.
\end{align}
Here, the time is expressed in units such that the spatial circle on which the boundary CFT lives has unit radius and $\varepsilon$ is a holographic regulator to be introduced in \eqref{eqn:holographic_regulator}.

\begin{figure}[t]
\includegraphics[width=12cm]{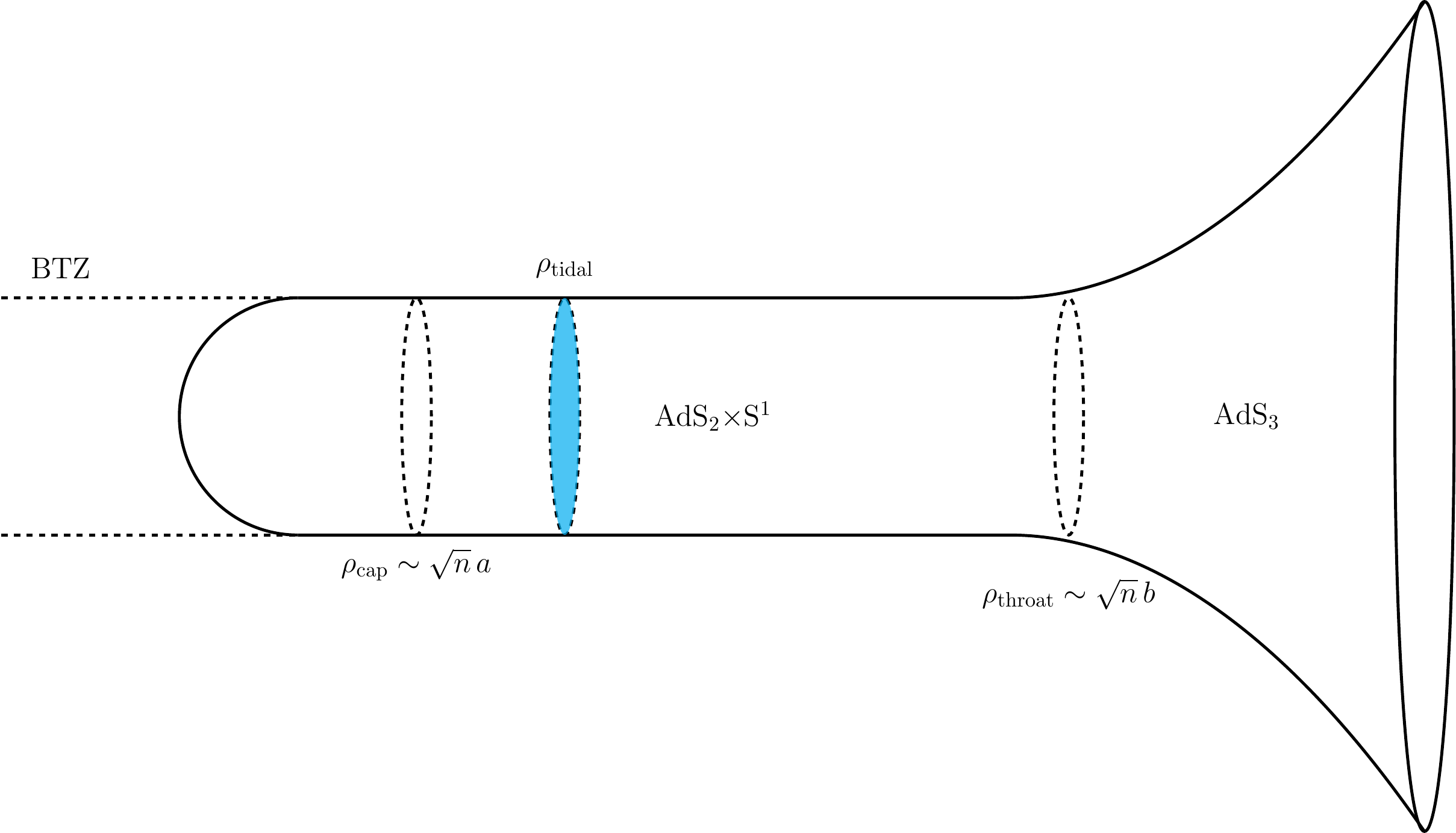}
\centering
\caption{Schematic representation of a spatial slice of the superstrata geometry described by the metric \eqref{superstrata}. Each point in the picture corresponds to an $S^3$ in the six-dimensional geometry. Coming in from infinity, there is an asymptotic $AdS_3\times S^3$ region, followed by an $AdS_2 \times S^1\times S^3$ throat region which then ends in a smooth cap region. For geodesics falling in from a large radius in the asymptotic region, the tidal forces are in danger of invalidating the geodesic approximation at a radius $\rho_{tidal}$ in the throat region.}
\label{fig:microstate_cap}
\end{figure}

Next, in section \ref{sec:microstate}, we turn to the microstate geometries studied in \cite{Bena:2016ypk,Bena:2019azk}. We focus on the $(1,0,n)$-superstrata whose salient geometric features are described in section~\ref{sec:superstrata}. They are constructed from a BPS configuration of $N_1$ D1-branes and $N_5$ D5-branes. These geometries have an asymptotic $AdS_3 \times S^3 \times T^4$ region. They closely approximate an $S^3 \times T^4$ trivially fibred over an extremal BTZ outside of a cap region. In the cap region, the fibration becomes non-trivial and the $S^3$ pinches off in a smooth way. This allows the geometry to smoothly end outside of the would-be horizon of extremal BTZ. This geometry can be pictured by considering a spatial slice like that of figure~\ref{fig:microstate_cap}. In extremal BTZ, this spatial slice is comprised of a near-horizon $AdS_2 \times S^1$ throat of infinite proper length attached to the asymptotic $AdS_3$ region. These microstates also exhibit a long throat attached to an asymptotic region, but this throat is capped off at a finite proper distance, before the would-be horizon is reached.
These geometries are dual to a class of states in a two-dimensional CFT with central charge $c=6 N_1 N_5$ \cite{Bena:2016ypk}.

In section \ref{sec:OTOC_superstrata} we describe the first steps towards computing the OTOC within the geodesic approximation in these microstate geometries. First, we study null geodesics with zero angular momentum in order to understand what regions of the geometry are probed by the OTOC as a function of the time separation between the operator insertions on the boundary. We find that for early times the interaction happens in the region of the geometry well described by extremal BTZ and so we expect the OTOC to be well described by the computations of section \ref{section:BTZ}. At a time scale
\begin{align}
    \label{eqn:t_cap-intro}
t_{cap}\simeq \frac{N_1 N_5}{T_R}\,,
\end{align}
the OTOC starts to probe the cap region of the geometry and so we expect it to strongly deviate from the extremal BTZ answer.
There are two main effects that will cause it to deviate: the blueshift in the center-of-mass energy will stop increasing since the geodesics cannot fall any further into the throat and the shock wave controlling the gravitational interaction between the scattered geodesics will be modified by the presence of the cap. These effects are discussed further in section \ref{sec:OTOC_superstrata}.

Another effect can invalidate the geodesic approximation before either of these effects of the cap manifest themselves. In \cite{Tyukov:2017uig}, it was found that the tidal forces in the throat region become Planckian well before the cap region. These tidal forces have been subsequently studied in a number of works including \cite{Bena:2018mpb,Bena:2020iyw,Martinec:2020cml}. The geodesic approximation requires that the volume expansion of a particular congruence of geodesics be much smaller than the mass of the particle, which will generically be violated in a region of large tidal forces. These tidal forces become important, and are therefore in danger of invalidating the geodesic approximation, at a time scale
\begin{align}
\label{eqn:t_tidal-intro}
t_{tidal} \simeq \sqrt{
    \frac{ \pi T_R}{ \pi^2 T_R^2+1 }
    \min(h_V,h_W) \varepsilon  N_1 N_5
    }\,.
\end{align}
 We have not tried to compute the shock wave produced by these geodesics in the cap region, since in any case the geodesic approximation does not hold in that region. Instead, the exact bulk-to-boundary propagators would need to be combined with the bulk-to-bulk graviton propagators to access the cap region, a computation we leave for future work.

The most relevant question for our purposes is whether the effects of the cap appear before or after the scrambling time. Indeed, given \eqref{eq:ts-intro} with the appropriate central charge, we find that $t_s \ll t_{tidal}$ for large black holes in the semi-classical limit as long as the right-moving temperature is not too high
\begin{align}
    T_R \ll \min(h_V,h_W)
    \left(\frac{h_V^2 h_W^2 N_1N_5}{\varepsilon} \right)^{\frac13}\,.
\end{align}
As long as this condition holds, we expect the commutator squared to stop growing well before the interaction region reaches the part of the superstrata geometry where it deviates from extremal BTZ.
In this case, we do not expect the details of the cap to affect the scrambling behaviour. They only come in far into the tail of the decay of the OTOC in the details of how the commutator squared saturates. However, for sufficiently high right-moving temperature, there does seem to be a regime where the effects of the cap will be felt before the scrambling time. We are not aware of any limit on the parameter $n$ appearing in the superstrata solutions, which means that such large temperatures are allowed. This region of large temperature would be an interesting regime to probe more precisely with a computation that goes beyond the geodesic approximation so that it can take into account the effects of the cap.

In section \ref{section:discussion}, we collect a number of open problems and directions for future work.

\paragraph{Conventions.} We work in units such that the AdS length $\ell_{AdS}=1$. The time coordinate in terms of which we express the various time scales agrees with the time coordinate of the dual CFT on a spatial circle with unit radius.

\section{Geodesic approximation to the OTOC}\label{section:geodesic}

We consider an asymptotically AdS background spacetime on which two massive Klein-Gordon real scalar fields $\phi_V$ and $\phi_W$ propagate. We are interested in computing the out-of-time-order correlator
\begin{equation}
\label{OTOC}
\text{OTOC}\equiv \langle \psi | \phi_V(X_1) \phi_W(X_2) \phi_V(X_3) \phi_W(X_4) |\psi \rangle,
\end{equation}
where the insertion points $X_2, X_4$ lie in the future of $X_1, X_3$ or are spacelike-separated from them. The state $|\psi\rangle$ corresponds to the background geometry on which the scalar fields propagate. All insertion points are also taken to lie asymptotically close to the spacetime conformal boundary in order to reproduce the OTOC of a dual conformal field theory. A method based on the geodesic approximation has been developed in a previous paper \cite{Balasubramanian:2019stt},
which may be viewed as a position-space version of the one originally presented by Shenker and Stanford~\cite{Shenker:2014cwa}.
It was similarly constructed as the overlap
\begin{equation}
\label{eq:overlap}
\text{OTOC}=\langle \textsl{out}|\textsl{in} \rangle,
\end{equation}
between the in- and out-states
\begin{equation}
|\textsl{in} \rangle\equiv \phi_V(X_3)\phi_W(X_4)| \psi\rangle, \qquad |\textsl{out} \rangle\equiv \phi_W(X_2)\phi_V(X_1)| \psi\rangle.
\end{equation}
The operator $\phi_W(X_4)$ used to create the in-state is represented on an early time slice $\Sigma_-$ by free propagation backward in time using the advanced propagator. In the same way, the operator $\phi_V(X_1)$ used to create the out-state is represented on a late time slice $\Sigma_+$ by free propagation forward in time using the retarded propagator. Note that the choice of these time slices is completely arbitrary and does not affect the end result. 
\begin{figure}[H] 
	\centering
	\subfloat[]{{\includegraphics[scale=0.42]{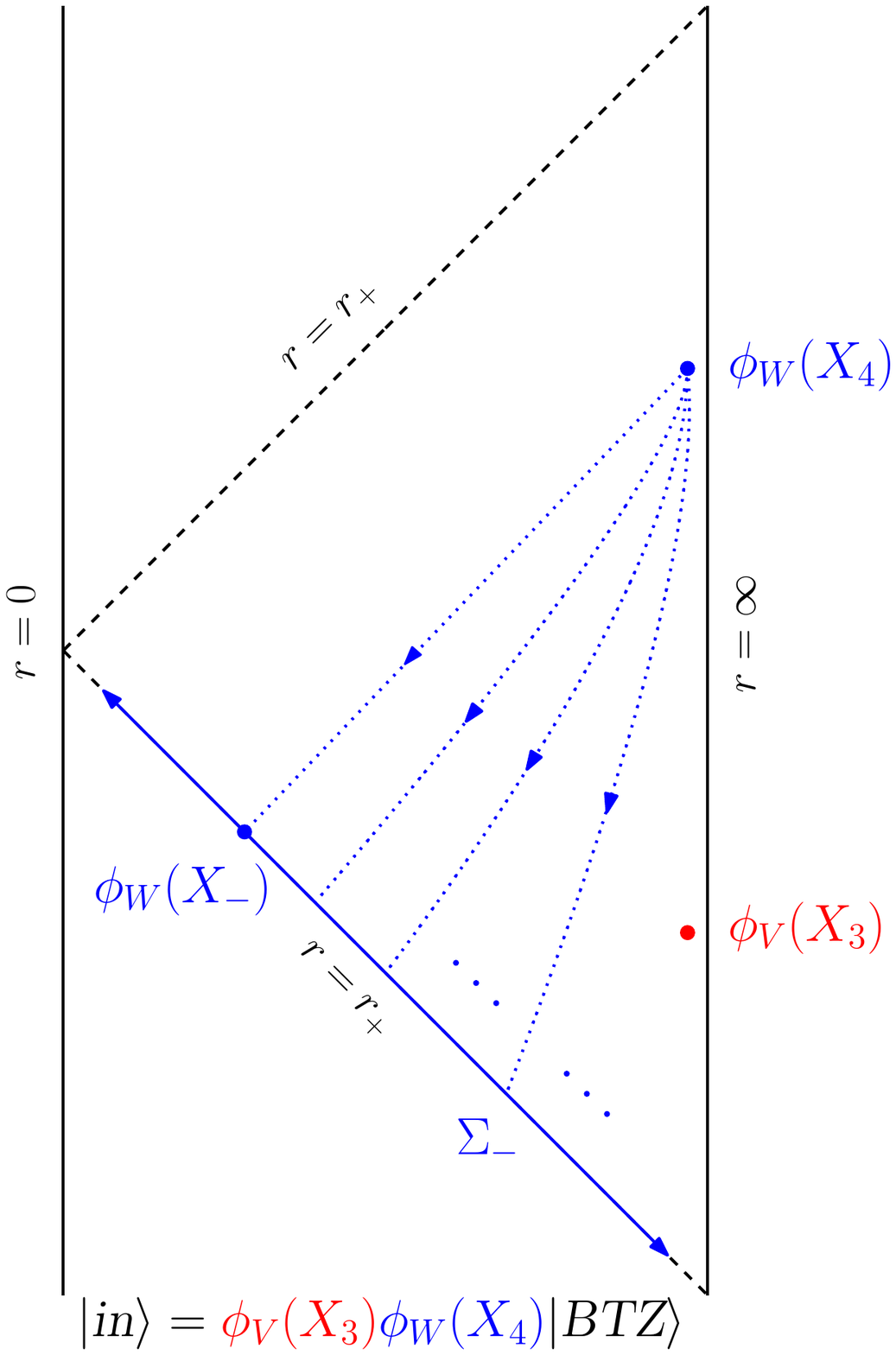} }}    \qquad
    \subfloat[]{{\includegraphics[scale=0.42]{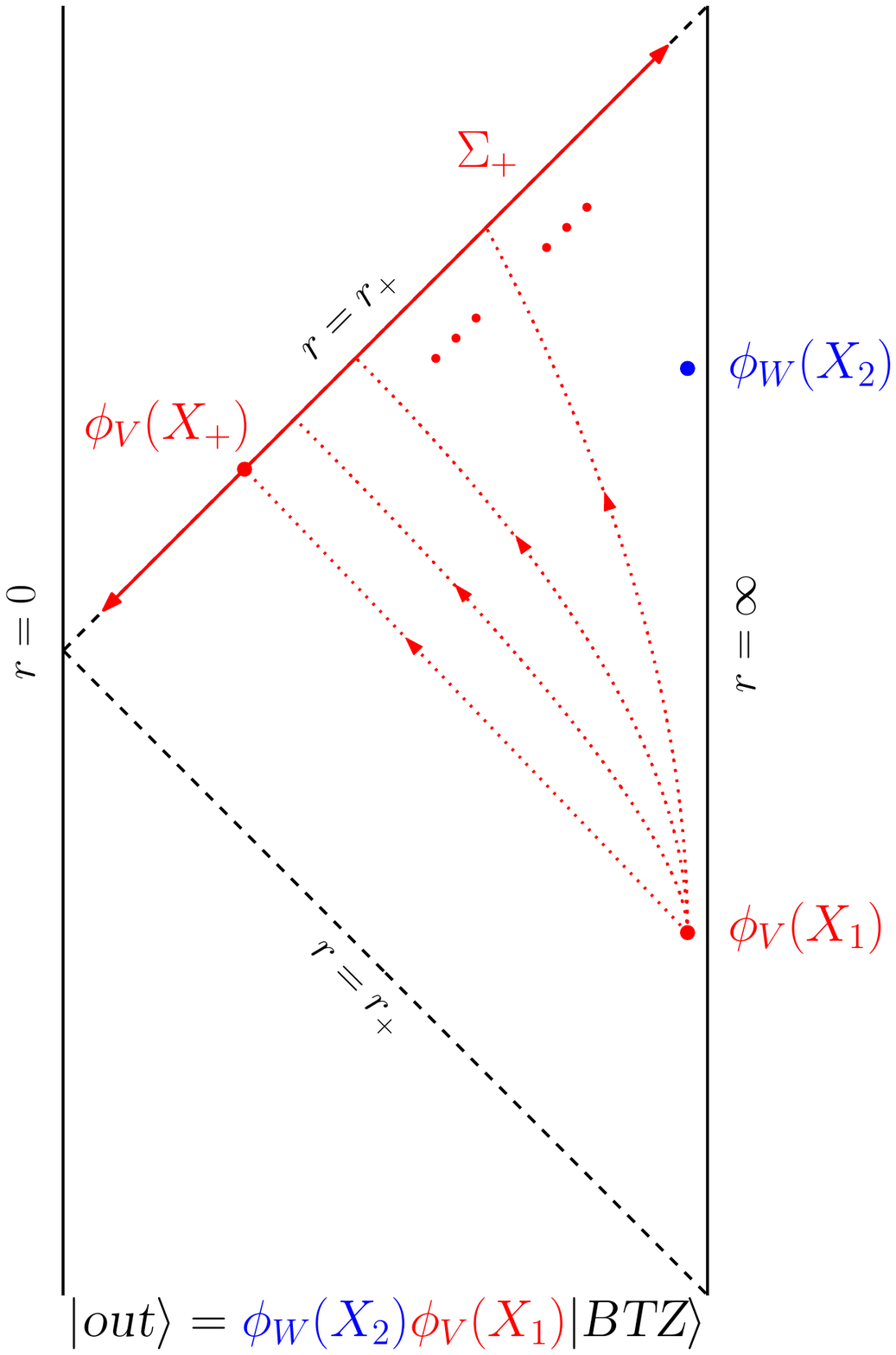} }}\\    \subfloat[]{{\includegraphics[scale=0.42]{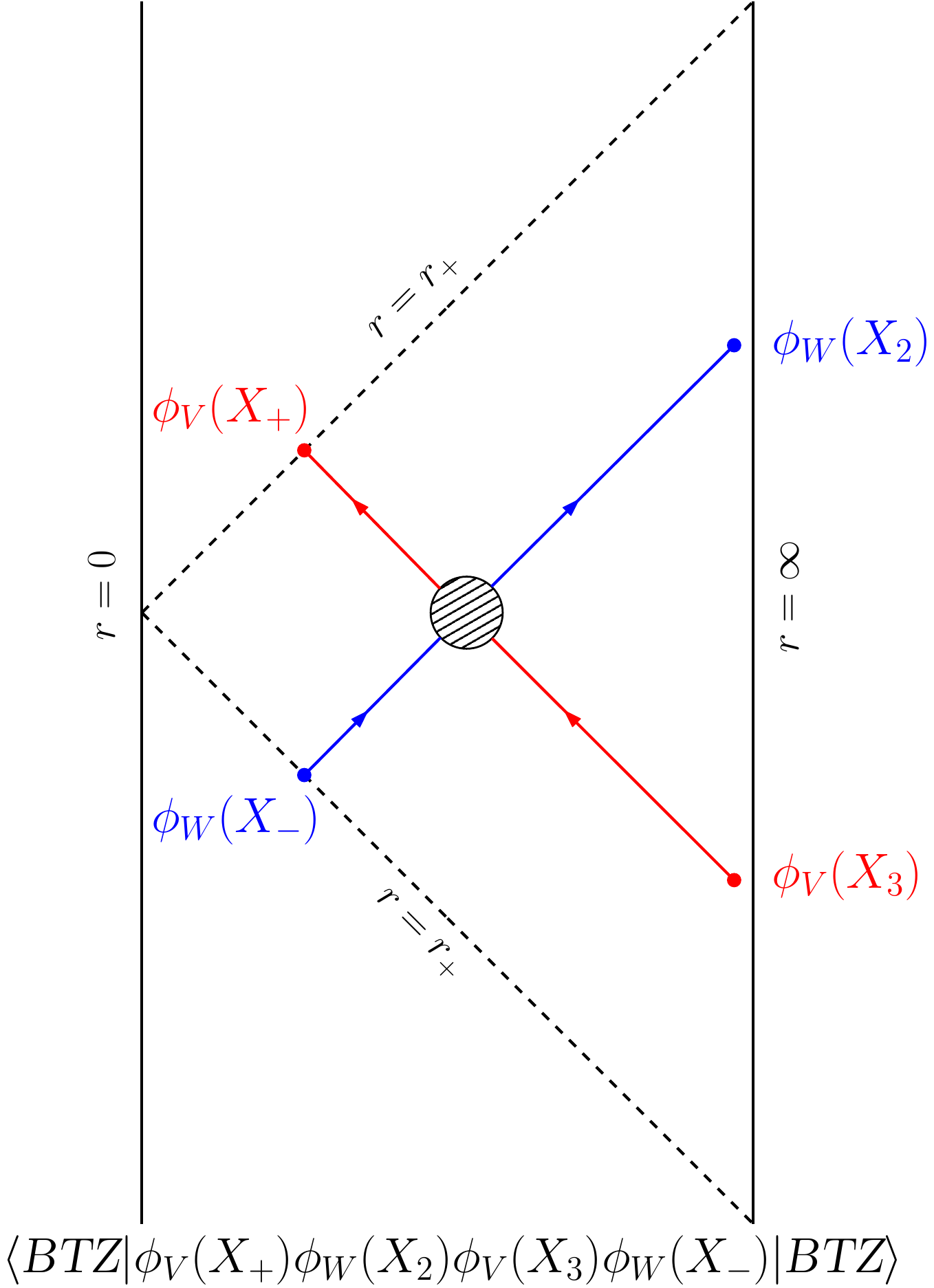} }}	\caption{Steps involved in the derivation of formula \eqref{overlap} for the OTOC written as the state overlap $\langle \textsl{out}|\textsl{in} \rangle$, illustrated in the case where the background geometry is an extremal BTZ black hole, i.e.~$|\psi\rangle =|\textsl{BTZ} \rangle$. (a)-(b) The operator $\phi_W(X_4)$ ($\phi_V(X_1)$) used to create the in-state (out-state) may be represented on the early (late) time slice $\Sigma_-$ ($\Sigma_+$) by free propagation backward (forward) in time. The slices $\Sigma_-$ and $\Sigma_+$ are chosen to coincide with the past and future horizons, respectively. (c) The state overlap $\langle \textsl{out}|\textsl{in} \rangle$ reduces to \textit{time-ordered} transition amplitudes involving all points $X_- \in \Sigma_- \cap \, J^-(X_4)$ and $X_+ \in \Sigma_+ \cap \, J^+(X_1)$. }
\label{fig:overlap}
\end{figure}
\noindent
The overlap \eqref{eq:overlap} then equals the \textit{time-ordered} transition amplitude
\begin{equation}
\label{eq:time-ordered}
\langle \psi| \phi_V(X_+) \phi_W(X_2) \phi_V(X_3) \phi_W(X_-)|\psi \rangle \sim e^{i\delta},
\end{equation}
convoluted with boundary-bulk propagators encoding the backward and forward propagation in time described above. 
Within the geodesic approximation $m_V, m_W \gg 1$ and the high-energy (eikonal) regime $G_N s\lesssim 1$, where $s$ is the center-of-mass energy of the corresponding 2-to-2 scattering, the time-ordered amplitude \eqref{eq:time-ordered} reduces to a simple phase $e^{i\delta}$ which we describe below in more detail.
This whole construction, originally presented in \cite{Balasubramanian:2019stt}, is illustrated for the case of an extremal BTZ background geometry in figure~\ref{fig:overlap}. For a more detailed description of the extremal BTZ geometry, we refer the reader to section~\ref{section:BTZ}, where we apply the general method presented here to this particular background spacetime.
All in all, this construction yields the formula
\begin{align}
\nonumber
\label{overlap}
\text{OTOC}
&=-4m_W m_V \int_{\Sigma_- \cap\, J^-(X_4)} d\Sigma \cdot k_W \ \Psi_W(X_2,X_-) \Psi_W(X_4,X_-)^*\\
&\hspace{0.5cm} \times \int_{\Sigma_+ \cap\, J^+(X_1)} d\Sigma \cdot k_V \ \Psi_V(X_+,X_3) \Psi_V(X_+,X_1)^*\  e^{i\delta},
\end{align}where $J^+(X_1)$ ($J^-(X_4)$) denotes the causal future (past) of the insertion point $X_1$ ($X_4$). 

In the geodesic approximation which we consider, boundary-bulk propagators appearing in the above formula are given by
\begin{equation}
\label{eq:WKB-propagator}
\Psi(X,Y)\equiv \langle \psi| \phi(X) \phi(Y)|\psi \rangle = A(X,Y) e^{i m S(X,Y)},
\end{equation}
where the point $X$ is assumed to lie in the causal future of the point $Y$, and with the phase given by the action of a timelike geodesic going from $Y$ to $X$ with velocity~$k$,
\begin{equation}
\label{eq:WKB-S}
S(X,Y)=\int_{Y}^X dx \cdot k.
\end{equation}
According to \eqref{overlap}, each point $X_-$ ($X_+$) within the time slice $\Sigma_-$ ($\Sigma_+$) that can be connected by a timelike geodesic to the future (past) insertion points $X_2, X_4$ ($X_1,X_3$) must be considered. Finally, the \textit{eikonal phase shift} $\delta$ encodes the gravitational interaction between the outgoing geodesic going from $X_-$ to $X_2$ and the ingoing geodesic going from $X_3$ to $X_+$. The stress-energy tensor $T_{\mu\nu}^V$ of an ingoing $V$ particle is the source of a gravitational field $h_{\mu\nu}^V$ that propagates and interacts with the outgoing geodesic, and conversely. For the eikonal phase shift, this yields the formula  \cite{Balasubramanian:2019stt}
\begin{equation}
\label{eikonal general}
\delta=\frac{1}{4} \int \left(h^V_{\mu\nu} T_W^{\mu\nu}+h^W_{\mu\nu} T_V^{\mu\nu}\right)+O(G_N^2).
\end{equation}
If it were not for the eikonal phase factor, the OTOC \eqref{overlap} would simply factorize into the product of two boundary propagators,
\begin{subequations}
\label{propagators}
\begin{align}
\label{propagator_1}
\langle \psi| \phi_V(X_1) \phi_V(X_3)|\psi \rangle&=2m_V \int_{\Sigma_+ \cap \, J^+(X_1)} d\Sigma \cdot k_V\ \Psi_V(X_+,X_3) \Psi_V(X_+,X_1)^*,\\
\label{propagator_2}
\langle \psi| \phi_W(X_2) \phi_W(X_4)|\psi \rangle&=-2m_W \int_{\Sigma_- \cap \, J^-(X_4)} d\Sigma \cdot k_W\ \Psi_W(X_2,X_-) \Psi_W(X_4,X_-)^*.
\end{align}
\end{subequations}

\paragraph{Stress tensor and shock wave of a particle.} The stress-energy tensor of a massive particle with trajectory $x^\mu(\tau)$ and velocity $k^\mu(\tau)$ may be conveniently  written \cite{Balasubramanian:2019stt}
\begin{equation}
\label{stress-tensor-general}
T_{\mu\nu}=\frac{m}{\sqrt{-g}} \left(\frac{dx^0}{d\tau}\right)^{-1} k_\mu k_\nu\, \delta(x^1-x^1(\tau))...\,  \delta(x^d-x^d(\tau))\Big|_{x^0=x^0(\tau)}.
\end{equation}
It sources a gravitational field which one may compute by solving the linearized Einstein's equations
\begin{equation}
D_{\text{lin}}\ h_{\mu\nu} =8\pi G_N\ T_{\mu\nu},
\end{equation}
where $D_{\text{lin}}$ is a differential operator whose definition involves the background geometry. For a single particle source with arbitrarily high energy, $h_{\mu\nu}$ is known as a gravitational \textit{shock wave}. Although it is found by solving the above linearized Einstein's equations, it is usually also a nonlinear solution \cite{Sfetsos:1994xa}.

\paragraph{Ultraviolet regulators.} We further restrict our attention to a configuration of infinitesimally separated boundary insertion points of the form
\begin{align}
\label{UV-regulator}
X_3^\mu=X_1^\mu-\epsilon_V\, \xi^\mu, \qquad X_4^\mu=X_2^\mu-\epsilon_W\, \xi^\mu,
\end{align}
where $\xi$ is a future-directed timelike vector of our choice that is tangent to the conformal boundary.\footnote{More precisely, $X_3=\exp{(-\epsilon_V \xi)}$, where $\exp$ is the exponential map at the point $X_1$.} The role of $\epsilon_V, \epsilon_W$ is to avoid UV divergences due to the insertion of operators at the same boundary points. When working in the limit of small UV regulators $\epsilon_W, \epsilon_V \ll 1$, the WKB phases in the overlap formula \eqref{overlap} simply differ by
\begin{subequations}
\begin{align}
S_V(X_+,X_3)&=S_V(X_+,X_1)+\epsilon_V\ \xi \cdot k_V(X_1)+\mathcal{O}(\epsilon_V^2),\\
S_W(X_4,X_-)&=S_W(X_2,X_-)-\epsilon_W\ \xi \cdot k_W(X_4)+\mathcal{O}(\epsilon_W^2),
\end{align}
\end{subequations}
such that
\begin{subequations}
\begin{align}
\label{phaseV}
\Psi_V(X_+,X_3) \Psi_V(X_+,X_1)^*&=\left|A_V(X_+,X_1)\right|^2 e^{im_V \epsilon_V \xi \cdot k_V(X_1)},\\
\label{phaseW}
\Psi_W(X_2,X_-) \Psi_W(X_4,X_-)^*&=\left|A_W(X_4,X_-)\right|^2 e^{im_W \epsilon_W \xi \cdot k_W(X_4)}.
\end{align}
\end{subequations}
Hence, the OTOC \eqref{overlap} simplifies to
\begin{align}
\label{overlap_BTZ}
\text{OTOC}
&=-4m_W m_V \int_{\Sigma_- \cap\, J^-(X_4)} d\Sigma \cdot k_W\ \left|A_W(X_4,X_-)\right|^2 e^{im_W \epsilon_W \xi \cdot k_W(X_4)}\\
\nonumber
&\hspace{0.5cm} \times \int_{\Sigma_+\cap \, J^+(X_1)} d\Sigma \cdot k_V\ \left|A_V(X_+,X_1)\right|^2 e^{im_V \epsilon_V \xi \cdot k_V(X_1)}\  e^{i\delta}.
\end{align}
It is customary to perform this integral by stationary phase approximation in the regime of large field masses $m_V, m_W$ \cite{Shenker:2014cwa,Balasubramanian:2019stt}.\footnote{As we explain in section~\ref{section:geodesics}, the velocities $k_{V,W}$ of particles inserted at a radial Schwarzschild coordinate $r=\varepsilon^{-1}$ scale like $O(\varepsilon^{-1})$ in the regime $\varepsilon \ll 1$. Thus, the stationary phase approximation to the integral \eqref{overlap_BTZ} holds within the regime $m_{V,W}\, \epsilon_{V,W} \gg \varepsilon$. Since $\epsilon_{V,W}$ and $\varepsilon$ both act as UV regulators, it is natural to consider them on the same footing, in which case the stationary phase approximation can be performed within the regime of large field masses $m_{V,W}\gg 1$.}

\paragraph{Early-time saddle point.} If one is only interested in the early-time Lyapunov growth, the eikonal phase can be neglected in determining the saddle point of the integral \eqref{overlap_BTZ}, and one simply has to extremize the initial component velocities $\xi \cdot k_V(X_1)$  and $\xi \cdot k_W(X_4)$ over the whole set of timelike geodesics connecting the boundary insertion points $X_1$ and $X_4$ to the time slices $\Sigma_-$ and $\Sigma_+$. We present a significant simplification in the determination of the dominant pair of geodesics compared to the method used in a previous publication \cite{Balasubramanian:2019stt}. This new method highlights that the arbitrary choice of time slices $\Sigma_-$ and $\Sigma_+$ does not affect the location of the saddle, since they do not enter the determination process at any point.  Instead of extremizing directly over the geodesic endpoints $X_- \in \Sigma_-$ and $X_+ \in \Sigma_+$, we equivalently extremize over their initial velocities $k_V$ and $k_W$, respectively. Thus, we need to determine the geodesic whose initial velocity $k^\mu$ is an extremum of the `energy' functional
\begin{equation}
E_\xi\equiv \xi \cdot k.
\end{equation}
Its variation with respect to the initial velocity is given by
\begin{equation}
\label{eq:delta-E}
\delta E_\xi=\xi \cdot \delta k,
\end{equation}
which is required to vanish for all allowed velocity variations $\delta k^\mu$. In fact, the only restriction on $\delta k^\mu$ comes from the timelike condition $k^2=-1$, whose variation yields
\begin{equation}
\label{eq:k-deltak}
k \cdot \delta k=0.
\end{equation}
From \eqref{eq:delta-E} and \eqref{eq:k-deltak}, we conclude that the initial velocity $k^\mu$ is a saddle point of $E_\xi$ if
\begin{equation}
\label{initial.velocity}
k^\mu \propto \xi^\mu,
\end{equation}
where we recall that $\xi$ is the future-directed timelike vector introduced in \eqref{UV-regulator}, and the normalization is easily found by imposing $k^2=-1$. The interpretation of this result is clear: the initial velocity of the geodesic inserted at $X_3\, (X_4)$ must point towards $X_1\, (X_2)$. Note that the same condition determines the saddle point of the two-point functions \eqref{propagators}. As a result, the \textit{normalized} OTOC takes a particularly simple form,
\begin{align}
\label{normalised OTOC}
\frac{\langle \phi_V(t_{in},x_-) \phi_W(t_{out},x_+) \phi_V(t_{in},x_-) \phi_W(t_{out},x_+) \rangle }{\langle \phi_V \phi_V \rangle \langle \phi_W \phi_W \rangle}\approx e^{i\delta}\Big|_{\text{saddle}},
\end{align}
where we have made implicit that all insertions happen arbitrarily close to the conformal boundary. The pair of geodesics corresponding to the dominant saddle in \eqref{normalised OTOC} is determined from the condition \eqref{initial.velocity} on their initial velocity at the boundary insertion points. In the limit where their insertion points are taken to the conformal boundary, these geodesics become approximately null. This is also precisely the regime in which the geodesic approximation to field propagation is most reliable and the gravitational field they create takes the form of shock waves \cite{Balasubramanian:2019stt}. However, note that the parameter $\varepsilon$ measuring how close to the conformal boundary the operators are inserted, to be introduced in section~\ref{section:BTZ}, acts as another UV regulator and cannot be taken to zero without introducing an appropriate renormalization scheme. We will not attempt to do this in the present paper.

The above approximation scheme breaks down in the regime where the eikonal phase shift itself significantly contributes to the determination of the dominant saddle of the integral \eqref{overlap}, i.e., when $\delta \gg 1$. Since $\delta$ is a growing function of time, which we describe in the next paragraph, this usually happens for late enough times. In particular, the quasi-normal decay of an OTOC (if it happens at all) lies within this late-time regime \cite{Shenker:2014cwa}. We refer the reader to appendix~\ref{app:oscillations} for further comments and details on the treatment needed in order to describe the quasi-normal decay of OTOCs.

\paragraph{Early-time Lyapunov growth.} Computation of OTOCs in the early-time regime through formula \eqref{normalised OTOC} instructs one to consider the stress tensor and gravitational field associated to one pair of highly energetic geodesics reaching the associated boundary insertion points, and to evaluate the eikonal phase shift \eqref{eikonal general} encoding their gravitational interaction. In the context of non-rotating BTZ black holes, it has been previously found that the latter scales with the center-of-mass energy $s$ of the 2-to-2 particle scattering \cite{Shenker:2014cwa,videomaldacena}
\begin{equation}
\delta \sim G_N s,
\end{equation}
where
\begin{equation}
\qquad s=-\left(m_V k_V+m_W k_W\right)^2 \approx -2m_Vm_W\,  k_V \cdot k_W, \qquad k_W^2=k_V^2=-1.
\end{equation}
The exponential Lyapunov growth originates from the exponential blueshift experienced by these particles in the neighborhood of a black hole. Indeed, one has
\begin{equation}
k_V(t_*) \sim e^{\kappa (t_*-t_{in})}\ k_V(t_{in}), \qquad k_W(t_*) \sim e^{\kappa (t_{out}-t_*)}\ k_W(t_{out}),
\end{equation}
where $t_*$ is the time of interaction and $\kappa$ is the surface gravity of the black hole background. Hence, one generically finds
\begin{equation}
\delta \sim G_N s \sim G_N m_W m_V e^{\kappa (t_{out}-t_{in})},
\end{equation}
which yields an exponential growth in the commutator squared, with Lyapunov exponent
\begin{equation}
\label{eq:Lyapunov-estimate}
\lambda_L=\kappa=\frac{2\pi}{\beta}.
\end{equation}
This simple reasoning is useful to estimate the exponential Lyapunov growth, but is in no way rigorous nor accurate. As an example, it has been shown that \eqref{eq:Lyapunov-estimate} only holds on average in the context of non-maximally rotating BTZ black holes \cite{Mezei:2019dfv}.

The above description does not apply to states at zero temperature $\beta \to \infty$, however. In the case of empty AdS dual to the CFT ground state, it has been shown that the eikonal phase $\delta$ grows quadratically with time \cite{Balasubramanian:2019stt}, finding agreement with earlier results obtained by CFT techniques \cite{Roberts:2014ifa}. This sort of polynomial growth has been associated to a form of \textit{slow scrambling}, in contrast to fast scrambling in case of exponential growth. In this paper we focus on extremal black holes to which a zero temperature is also associated. We study the case of a maximally rotating BTZ black hole in section~\ref{section:BTZ}, and show that the growth in time of the eikonal phase alternates between quadratic and exponential -- with Lyapunov exponent associated to the nonzero `right-moving' temperature. On average, the growth is cubic so that the scrambling may be qualified as slow. We will phrase this latter result in terms of the center-of-mass energy of the corresponding 2-to-2 particle scattering together with the topology of the black hole. We turn to superstratum microstate geometries in section~\ref{section:microstate}, and point to the various effects that potentially distinguish the behaviour of the eikonal phase and OTOC, compared to the case of an extremal BTZ geometry.

\section{OTOC in extremal BTZ}\label{section:BTZ}

We now focus on the OTOC computation in the particular case of maximally rotating BTZ black holes. The latter being extremal and therefore having zero temperature, we expect  additional subtleties compared to the computation of OTOCs in non-extremal BTZ black holes \cite{Shenker:2014cwa,Jahnke:2019gxr,Mezei:2019dfv,Balasubramanian:2019stt}.

The exterior region of extremal BTZ is commonly described using Schwarzschild coordinates $(t,r,\varphi)$ with metric\footnote{Useful formulae may be found in \cite{Gralla:2019isj}.}
\begin{equation}
\label{static-extremal-metric}
ds^2=\ell_{AdS}^2\left[-\left(r^2-2 r_+^2 \right) dt^2+\frac{r^2\ dr^2}{\left(r^2-r_+^2 \right)^2}-2 r_+^2\ dt d\varphi +r^2\ d\varphi^2\right],
\end{equation}
where the angular coordinate is periodically identified, $\varphi \sim \varphi+2\pi$. The black hole horizon lies at $r=r_+$ while a timelike singularity lies at $r=0$. The AdS conformal boundary lies at $r \to \infty$, has cylinder topology and is covered by the coordinate system $(t,\varphi)$. The Penrose diagram of extremal BTZ is displayed in figure~\ref{fig:overlap}. See \cite{Carlip:1995qv,Compere:2018aar} for thorough reviews of three-dimensional BTZ black holes. In what follows, we will display time scales in terms of the time coordinate $t$, which is also the time coordinate of the dual CFT with a spatial circle of unit radius.

Being extremal, this black hole has zero Bekenstein-Hawking temperature. However, because it corresponds to a rotating ensemble, one can associate distinct temperatures to right- and left-moving modes,
\begin{equation}\label{eq:temp}
T_L=0, \qquad T_R=\frac{r_+}{\pi}.
\end{equation}
In particular, right-movers are at nonzero temperature. Extrapolating earlier results found in the case of non-maximally rotating BTZ black holes \cite{Mezei:2019dfv,Jahnke:2019gxr}, we can expect that the OTOC of scalar operators alternates between a polynomial growth associated to a zero left-moving temperature $T_L$ and an exponential growth associated to a nonzero right-moving temperature $T_R$. We will show in section~\ref{subsection: eikonal phase} that this is indeed the case.

We specify the coordinates of the boundary insertion points $X_1$ and $X_2$ of the OTOC as follows:
\begin{align}
t_1&\equiv t_{in}, \qquad \varphi_1\equiv \varphi_{in}, \qquad r_1=\varepsilon^{-1}, \label{eqn:holographic_regulator}\\
t_2&\equiv t_{out}, \qquad \varphi_2\equiv \varphi_{out}, \qquad r_2=\varepsilon^{-1}.
\end{align}
Here, we consider $\varepsilon \ll 1$ as a holographic regulator measuring how close to the conformal boundary operators are inserted. The specification of the other two boundary insertion points $X_3$ and $X_4$ is made through a choice of point-splitting regulator $\xi$ of the type \eqref{UV-regulator}, which we make in such a way that both ingoing and outgoing geodesics connecting the above insertion points have zero angular momentum. This is always possible to achieve, and we leave the expression of $\xi$ implicit. Following the geodesic approximation described in section~\ref{section:geodesic},
the computation of the OTOC at early times $\delta \lesssim 1$ (see section~\ref{section:geodesic})
amounts to the evaluation of the eikonal phase factor \eqref{eikonal general} encoding the gravitational interaction of the two geodesics. As we will show, the energy $E$ of the associated particles scales like $E\sim \varepsilon^{-1}$ such that, in the limit $\varepsilon \to 0$, they follow null trajectories.

\subsection{Highly energetic particles}
\label{section:geodesics}
Evaluation of the OTOC through the geodesic approximation scheme presented in section~\ref{section:geodesic} requires one to consider the ingoing timelike geodesic connecting the  insertion point $(\varepsilon^{-1},t_{in},\varphi_{in})$, with initial velocity $k^\mu$ proportional to the point-splitting regulator $\xi^\mu$ as shown in \eqref{initial.velocity}. The outgoing timelike geodesic connecting the insertion point $(\varepsilon^{-1},t_{out},\varphi_{out})$ has to be considered similarly. We start by showing that at leading order in $\varepsilon \ll 1$, we can switch to a description in terms of null geodesics. We then give the expressions of the stress tensor and gravitational shock wave associated to each one of these null geodesics, which will be needed in section~\ref{subsection: eikonal phase} in order to compute the eikonal phase and OTOC.

\paragraph{Timelike geodesics.}
A timelike geodesic with velocity $k^\mu=\dot{x}^\mu=d x^\mu/d\tau$ has conserved energy\footnote{Following the choice of normalization $k^2=-1$, $E$ and $L$ are the conserved energy and angular momentum \textit{per unit mass}.} $E$ and angular momentum $L$ associated to the Killing vectors $\partial_t$ and $\partial_\varphi$ of the extremal BTZ metric \eqref{static-extremal-metric},
\begin{equation}
E=-(\partial_t)^\mu k_\mu, \qquad L=(\partial_\varphi)^\mu k_\mu, \qquad k^2=-1.
\end{equation}
In terms of these conserved quantities, the radial velocity of a timelike geodesic satisfies \cite{Cruz:1994ir}
\begin{equation}
\label{eq:rdot-timelike}
r^2\dot{r}^2=-(r^2-r_+^2)^2+ \left(E^2-L^2\right)r^2+2 \left(L^2-EL\right)r_+^2.
\end{equation}
The choice of point-splitting regulator $\xi^\mu$ determines the velocity $k^\mu$ of the geodesics at the insertion points. Since $\xi^\mu$ is tangent to the (cutoff) boundary, the latter necessarily has vanishing radial component, $\dot{r}=0$. In addition, we choose the orientation of $\xi^\mu$ in such a way that these geodesics also have zero angular momentum $L=0$. Plugging these requirements into \eqref{eq:rdot-timelike}, we find the value of the energy,
\begin{equation}
\label{eqn:energy}
E=\varepsilon^{-1}\left(1-\varepsilon^2 r_+^2\right).
\end{equation}
In the limit $\varepsilon \to 0$ where the insertion points are taken to the conformal boundary, the energy of these geodesics simply diverges and their trajectories coincide with those of the corresponding null geodesics; see also \cite{Balasubramanian:2019stt}. In the following, we work at leading order in $\varepsilon \ll 1$, at which we can simply approximate the trajectories of the highly energetic particles of interest by null geodesics with energy $E=\varepsilon^{-1}$ and angular momentum $L=0$. We switch to this leading order approximation in what follows.

\paragraph{Null geodesics and shock waves.}
We thus restrict our attention to null geodesics. In terms of the null velocity $k^\mu$, the conserved energy and angular momentum are
\begin{equation}
\label{eq:conserved quantities}
E=-(\partial_t)^\mu k_\mu, \qquad L=(\partial_\varphi)^\mu k_\mu, \qquad k^2=0.
\end{equation}
The null geodesic equations are then given by \cite{Cruz:1994ir}
\begin{subequations}
\begin{align}
\dot{t} &= \frac{E r^2-L r_+^2}{\left(r^2-r_+^2\right)^2}, \label{eq: dt} \\
\dot{\varphi} &= \frac{E r_+^2+L (r^2-2r_+^2)}{\left(r^2-r_+^2\right)^2}, \label{eq: dx} \\
r^2\dot{r}^2 &=\left(E^2-L^2\right)r^2+2 \left(L^2-EL\right)r_+^2.
\label{eq: dr}
\end{align}
\end{subequations}

To describe the ingoing geodesic, it is convenient to define retarded coordinates $(r,v,\phi)$ through
\begin{subequations}
\label{eq:retarded_coordinates}
\begin{align}
\label{eqn:retarded_coordinates 1}
t&=v+\frac{r}{2\left(r^2-r_+^2\right)}-\frac{1}{4r_+} \ln \frac{r-r_+}{r+r_+},\\
\label{eqn:retarded_coordinates 2}
\varphi&=\phi+v+\frac{r}{2\left(r^2-r_+^2\right)}+\frac{1}{4r_+} \ln \frac{r-r_+}{r+r_+},
\end{align}
\end{subequations}
such that the extremal BTZ metric becomes
\begin{equation}
\label{eqn:infalling-coords}
ds^2=2 dr dv+2\left(r^2-r_+^2\right)d\phi dv+r^2 d\phi^2.
\end{equation}
In these coordinates, the coefficient of $d\phi dv$ vanishes at the horizon so that these coordinates give the co-rotating frame for infalling particles at the horizon.
In retarded coordinates, the ingoing geodesic of interest with $E=\varepsilon^{-1}$ and $L=0$ takes a particularly simple form. By differentiating equations \eqref{eqn:retarded_coordinates 1} and \eqref{eqn:retarded_coordinates 2} and using the negative root of \eqref{eq: dr} for $\dot{r}$, one finds that its velocity is purely radial,
\begin{equation}
k_V=-\varepsilon^{-1} \partial_r,
    \label{eq: momentum V}
\end{equation}
such that its trajectory is
\begin{equation}
v(r)=v_{in}=t_{in},
 \qquad  \phi(r)=\phi_{in}= \varphi_{in}-t_{in} \label{eq: trajectory in}.
\end{equation}
From \eqref{stress-tensor-general}, its stress tensor has only one non-trivial component,
\begin{equation}
T_V^{rr}=-\frac{m_V \varepsilon^{-1}}{\sqrt{-g}}\ \delta\left(v-v_{in}\right) \delta\left(\phi-\phi_{in}\right).
\label{eq: stress tensor ingoing}
\end{equation}
On the other hand, the velocity of the outgoing geodesic with $E=\varepsilon^{-1}$ and $L=0$ is
\begin{equation}
k_W=\varepsilon^{-1} \left(\frac{2r^2\partial_v}{\left(r^2-r_+^2\right)^2}+\partial_r-\frac{2\partial_\phi}{r^2-r_+^2} \right),
\label{eq: momentum W}
\end{equation}
while its trajectory can be found by integrating the velocity and can be parametrized by
\begin{subequations}
\label{eq: trajectory out}
\begin{align}
v(r)&=t_{out}-\frac{r}{r^2-r_+^2}+\frac{1}{2r_+} \ln \frac{r-r_+}{r+r_+},
\label{eq: trajectory out v}\\
\phi(r)&=\varphi_{out}-t_{out}-\frac{1}{r_+} \ln \frac{r-r_+}{r+r_+}.
\label{eq: trajectory out phi}
\end{align}
\end{subequations}
Retarded coordinates are well suited to compute the term $h^W_{\mu\nu} T^{\mu\nu}_V$ in the eikonal phase \eqref{eikonal general}, due to the simple form of the stress tensor \eqref{eq: stress tensor ingoing}. On the other hand, to compute the other term $h^V_{\mu\nu} T^{\mu\nu}_W$ in \eqref{eikonal general}, it is more convenient to turn to the advanced coordinate system $(r,u,\phi')$ defined through
\begin{subequations}
\label{advanced-coord}
\begin{align}
t&=u-\frac{r}{2\left(r^2-r_+^2\right)}+\frac{1}{4r_+} \ln \frac{r-r_+}{r+r_+},\\
\varphi&=\phi'+u-\frac{r}{2\left(r^2-r_+^2\right)}-\frac{1}{4r_+} \ln \frac{r-r_+}{r+r_+}.
\end{align}
\end{subequations}
In the following we focus on the $h^W_{\mu\nu} T^{\mu\nu}_V=h^W_{rr} T^{rr}_V$ contribution. We have checked that, as in previous work \cite{Shenker:2014cwa}, this is also equal to the $h^V_{\mu\nu} T^{\mu\nu}_W$ contribution.

The relevant component of the shock wave sourced by the outgoing particle is computed in appendix~\ref{appendix:shock wave} and is obtained from \eqref{eq: shock wave-outgoing compactified},
\begin{align}
\label{eq: shock wave-outgoing compactified main}
h^W_{rr} &=-\frac{4 \pi G_N m_W}{r_+^2 \varepsilon} \sum_{n\in \mathbb{Z}}  f\left(r_+\Delta v,r_+(\Delta v+\Delta \phi_n)\right)\, \delta \left(r- r_0(v,\phi) \right) \Theta(-\Delta \phi_n) \Theta(2\Delta v+\Delta \phi_n),
\end{align}
with
\begin{align}
r_0(v,\phi) &=\frac{1}{2\Delta v+ \Delta \phi_n}-r_+ \coth \left( r_+\Delta \phi_n  \right),
\end{align}
and
\begin{align}
\label{eq:f(t,x)}
f(t, x ) \equiv
\begin{cases}
(t+x)^2 \,, \quad & t+x < \sinh (t-x) \,,\\
\sinh^2 (t-x)\,, \quad & t+x \geq \sinh (t-x) \,,
\end{cases}
\end{align}
where $\Delta v = v_{out} - v$, $\Delta \phi_n = \phi_{out} - \phi + 2\pi n$ and $\Theta$ denotes the Heaviside function.
Note that, due to the $\phi$-direction being periodic, the shock wave includes the contribution of many `images' of the outgoing geodesic, appearing in \eqref{eq: shock wave-outgoing compactified main} through the summation over $n \in \mathbb{Z}$. The shock wave of a single image geodesic has support on a surface determined by the Dirac delta function and the Heaviside functions in \eqref{eq: shock wave-outgoing compactified main}, in such a way that a given bulk point ($v,r,\varphi$) lies at most on a finite number of shock wave images. Indeed, for fixed values of the coordinates, only a finite number of images satisfy $-2\Delta v \leq \Delta \phi_n \leq 0$.

\subsection{Eikonal phase}
    \label{subsection: eikonal phase}
\begin{figure}[th]    \centering
    \subfloat[]{{\includegraphics[width=7cm]{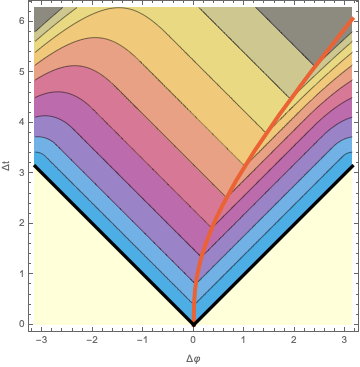} }}    \qquad
    \subfloat[]{{\includegraphics[width=7cm]{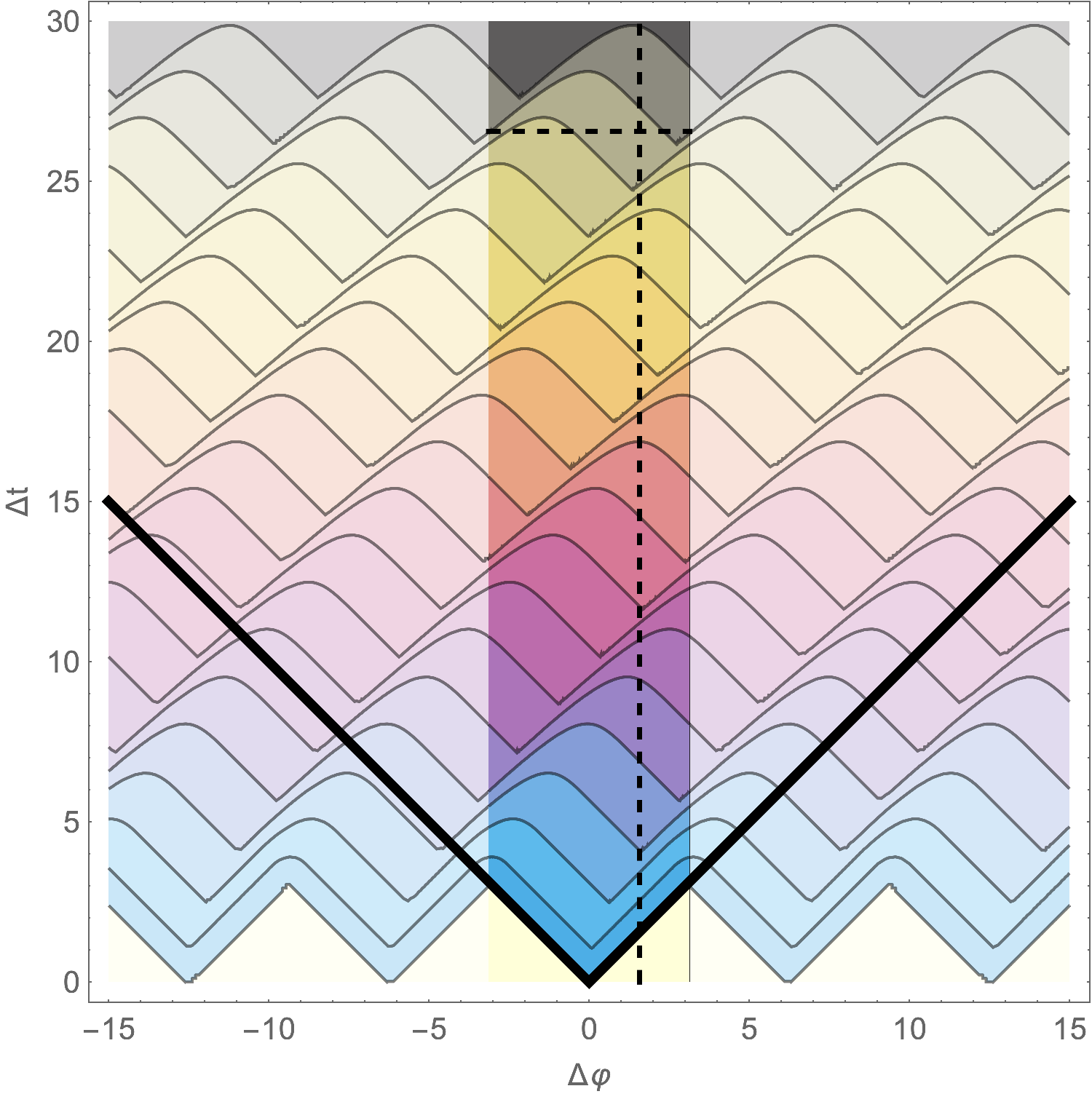} }}    \caption{Contour plots of the eikonal phase $\delta(\Delta \varphi,\Delta t)$ in extremal BTZ, for $r_+=1$, and where the $n^{th}$ contour level is given by a cubic function $(c n + d)^3$ with $c= 10^{-7}$ and $d= 10^{-7}$ in (a) and $(c n + d)^3$ with $c= 10^{-7}$ and $d= 15 \cdot 10^{-7}$ in (b). A cubic function would therefore have equidistant contour lines. (a) At small times, a single image contributes to the eikonal phase. The orange line characterizes the spatial separation for which $\delta$ is maximal at a fixed time separation, which corresponds to the boundary insertions for which the two geodesics cross with zero impact parameter in the bulk. On the right of this line, the growth as a function of time is exponential, while on the left there is power law growth. (b) The contour plot is extended for larger time separations, involving multiple images, while highlighting the physical domain $\Delta \varphi \in ]-\pi,\pi ]$. The behaviour of the eikonal phase along the vertical and horizontal dashed lines are shown in figures \ref{fig:OTOC_cubic} and \ref{fig: shock wave}, respectively.
    }
    \label{fig:contourplots}\end{figure}

With the analysis of the previous section, we are in a position to compute the eikonal phase \eqref{eikonal general}. Using \eqref{eq: stress tensor ingoing} and \eqref{eq: shock wave-outgoing compactified main}, we find
\begin{align}
    \delta &= \frac{1}{4} \int \sqrt{-g}\left( h^V_{\mu\nu}  T_W^{\mu\nu}+ h^W_{\mu\nu}  T_V^{\mu\nu}\right)=\frac{2 \pi G_N m_V m_W}{r_+^2\varepsilon^2}
    \sum_{\substack{n \in \mathbb{Z} \\ | \Delta \varphi_n | \leq \Delta t}}
      f(r_+ \Delta t,\, r_+ \Delta \varphi_n)\,,
    \label{eq: eik phase BTZ}
\end{align}
with  $\Delta t = t_{out} - t_{in}$ and $\Delta \varphi_n = \varphi_{out} - \varphi_{in} + 2 \pi n$.
Note that in terms of the boundary coordinates $(t,\varphi)$,  the condition  $ -2 \Delta v \leq \Delta \phi_n \leq 0$ on the sum translates to $| \Delta \varphi_n | \leq \Delta t$. Thus boundary causality determines which images should be included in the sum. Figure~\ref{fig:contourplots} displays the level sets of the eikonal phase. In figure~\ref{fig:contourplots} (a), we see the contribution of the first image. The orange line tracks the cusp in the eikonal phase, and corresponds to boundary insertions such that the dual particles collide in the bulk. The distinct accumulation of contour lines on each side of this orange line is due to the piecewise behaviour of the function $f(t,x)$ defined in \eqref{eq:f(t,x)}, which is quadratic on the left and exponential on the right. Figure~\ref{fig:contourplots} (b) shows the full result which involves a sum over images. The fictitious region from which `image particles' are emitted is represented in lighter colors, while the physical periodic region corresponding to $\Delta \varphi \in \left]-\pi,\pi \right]$ is brighter, with $\Delta \varphi = \varphi_{out}-\varphi_{in}$.

The time-dependence of the eikonal phase is easier to visualize from figure~\ref{fig:OTOC_cubic}, which corresponds to the dashed vertical cut of figure~\ref{fig:contourplots}. To properly understand the various features that appear, it is useful to further analyze the result \eqref{eq: eik phase BTZ}. Consider the sum appearing in \eqref{eq: eik phase BTZ}. It can be split into two elementary sums according to the two cases of the piecewise function,
\begin{align}
\label{eq:sum f}
 \sum_{\substack{n \in \mathbb{Z} \\ | \Delta \varphi_n | \leq \Delta t}}
     & f(r_+ \Delta t,\, r_+ \Delta \varphi_n) =\mathbb{I} + \mathbb{II}\,,
\end{align}
\begin{equation}
      \mathbb{I}=  \sum_{n= -\left\lfloor \frac{\Delta t +\Delta \varphi}{2\pi} \right\rfloor}^{ \left\lfloor n_* \right\rfloor}
      r_+^2 ( \Delta t + \Delta \varphi + 2 \pi n)^2\,, \qquad
      \mathbb{II}=\sum_{n= \left\lceil n_* \right\rceil}^{ \left\lfloor \frac{\Delta t -\Delta \varphi}{2\pi} \right\rfloor}
      \sinh^2 \left[r_+ (\Delta t - \Delta \varphi - 2 \pi n) \right]\,,
\end{equation}
where $n_*$ is the real number at which the transition between the two cases occurs, and satisfies
\begin{align}
    r_+ &(\Delta t + \Delta \varphi + 2 \pi n_*) = \sinh \left[r_+ (\Delta t - \Delta \varphi - 2 \pi n_*) \right]\,.
\end{align}
These sums are simple to evaluate, but the full expression is a bit involved so we will not write it out in full here. We will focus instead on the behaviour in the late time regime. The transition point $n_*$ can be approximated in the large $\Delta t \gg 1$ regime by,
\begin{align}
    \label{eq:nstar}
    2 \pi n_*= \Delta t -\Delta\varphi -\frac{\log (4 r_+ \Delta t) }{r_+} + O\left( \frac{\log (r_+ \Delta t)}{r_+^2 \Delta t} \right)\,.
\end{align}
Using this approximation along with the fact that $\lfloor x \rfloor = x - (x~\mathrm{mod}~1)$, we obtain
\begin{align}
      \delta =&
      \frac{4 \pi G_N m_V m_W }{\varepsilon^2}
           \Bigg\{  \frac{2 \Delta t^3}{3 \pi}
      -\frac{\Delta t^2 \log (4 r_+ \Delta t)}{\pi r_+} \label{eqn:sum_answer}\\
      &\,
      + \left( 1 -
      \frac{\left[ \left( \Delta t -\Delta \varphi -\frac{\log(4r_+ \Delta t)}{r_+} \right)~\mathrm{mod}~2\pi \right]}{\pi}
      +2\frac{e^{ 2 r_+ \big[ \left(\Delta t -\Delta \varphi - \frac{\log4 r_+ \Delta t }{r_+}\right)  ~\mathrm{mod}~2\pi \big]} }{ e^{4 \pi r_+}-1 }
      \right) \Delta t^2
      +O(\Delta t) \Bigg\}. \nonumber
\end{align}

\paragraph{Average slow scrambling.} The average growth of the eikonal phase for times $\Delta t \gg 1$ is well described by equation \eqref{eqn:sum_answer}. In particular, it is cubic up to subleading contributions. One of these subleading contributions is of the form $\left(\text{periodic function}\right)\times \Delta t^2$, which suggests a sawtooth-like behaviour around this average cubic growth. Indeed, we can observe in figure~\ref{fig:OTOC_cubic} that the average value of the eikonal phase follows the cubic orange dashed line, with sawtooth-modulation around this cubic growth. The scrambling time $\Delta t_s$ defined as the time at which $\delta \sim 1$, is inferred from this average cubic growth,
\begin{align}
\label{eqn:scrambling_time_3d}
   \Delta t_s &\simeq \left( \frac{3  \varepsilon^2}{8 G_N m_V m_W } \right)^{\frac13}=\left( \frac{c\, \varepsilon^2}{16 h_V h_W } \right)^{\frac13}\,.
\end{align}
The second equality gives the expression of the scrambling time in terms of CFT quantities, namely the central charge $c=\frac{3}{2G_N}$ and the conformal weights $2h_{V,W}\approx m_{V,W}\gg 1$ and reproduces \eqref{eq:ts-intro} from the introduction.

The average cubic growth could have been obtained in a much simpler fashion by considering the dependence of the center-of-mass energy of the scattered particles on the time separation of their boundary insertion points, together with the angular periodicity of the background spacetime leading to contributions from multiple image particles. Indeed, the center-of-mass energy $s$ associated to the scattering of the two null geodesics with momentum \eqref{eq: momentum V} and \eqref{eq: momentum W} and colliding at $r=r_*$ is proportional to
\begin{align}
k_V \cdot k_{W}= -\frac{2 r_*^2}{\varepsilon^2 (r_*^2-r_+^2)^2}.
\end{align}
From the trajectories of the geodesics given by \eqref{eq: trajectory in} and \eqref{eq: trajectory out}, one can relate $r_*$ to the time separation of the boundary points,
\begin{align}
    \Delta t = t_{out}-t_{in} = \frac{r_*}{r_*^2-r_+^2}-\frac{1}{2r_+} \ln \frac{r_*-r_+}{r_*+r_+}.
\end{align}
At late times, the geodesics scatter close to the horizon so that the separation is given at leading order by
\begin{align}
    \Delta t = \frac{1}{2(r_*-r_+)}+O\left( \ln \left(\frac{r_*}{r_+}-1 \right) \right),
\end{align}
such that the center-of-mass energy scales like
\begin{align}
    k_V \cdot k_W=\frac{\varepsilon^{-2}}{2(r_*-r_+)^2}+O\left(\left( \frac{r_*}{r_+}-1 \right)^{-1}\right)
    \approx 2 \varepsilon^{-2} \Delta t^2 \, .
\label{eqn:com_ext_BTZ}
\end{align}
Hence, the center-of-mass energy of the particle scattering grows quadratically with $\Delta t$. The discrepancy between this estimate and the cubic growth found from \eqref{eqn:sum_answer} is explained by the angular periodicity of the solution. Indeed, the number $n$ of image particles contributing to the sum \eqref{eq:sum f} actually grows linearly with time. Equivalently, the two physical scattered particles are seen to circle around the black hole and the number of times they meet grows linearly with $\Delta t$.  In summary, the average cubic growth of the eikonal phase $\delta$ may be inferred from the quadratic growth of the center-of-mass energy $s$ together with the linear growth in the number of particle images.

\paragraph{Sawtooth pattern.}
We would now like to understand the sawtooth pattern appearing on top of the average growth in figure \ref{fig:OTOC_cubic}. For this it is useful to analyze \eqref{eq:sum f} in more detail.
The two sums, $\mathbb{I}$ and $\mathbb{II}$, have very different behaviours. First, we note that for $\Delta t \gg1$ the range of $n$ in $\mathbb{I}$ scales as $\Delta t/\pi$ whereas in $\mathbb{II}$ it scales as $r_+^{-1} \log(4 r_+ \Delta t)$. Since the summand in $\mathbb{I}$ is a $O(\Delta t^2)$ polynomial, a sum over $\Delta t$ terms leads to a $\Delta t^3$ scaling. On the other hand, the summand in $\mathbb{II}$ is exponential and so its scaling is controlled by the largest term at $n=\lceil n_* \rceil$. The logarithmic term in $n_*$ leads to an overall $\Delta t^2$ growth for this sum,
\begin{align}
    \sinh^2 \left[ r_+(\Delta t - \Delta\varphi - 2 \pi \lceil n_* \rceil) \right] \sim 4 r_+^2 \Delta t^2 e^{-4 \pi r_+} e^{ 2 r_+ \big[ \left(\Delta t -\Delta \varphi - \frac{\log4 r_+ \Delta t }{r_+}\right)  ~\mathrm{mod}~2\pi \big]}  \,.
\end{align}
This explains why $\mathbb{I}$ dominates the overall value of the eikonal phase.

Let us now consider the sum $\mathbb{I}$ in more detail. On small time intervals, this sum has a fixed number of terms, each contributing an $O(\Delta t^2)$ growth, so it exhibits quadratic growth. An example of such quadratic growth is represented by the dotted green line in figure~\ref{fig:OTOC_cubic}. On larger time scales, the number of terms jumps discretely as additional images are included, and grows on average as $O(\Delta t)$. Thus the locally quadratic but overall cubic growth comes from a behaviour that can be schematically written as
\begin{align}
    \mathbb{I} \sim \lfloor \Delta t \rfloor \Delta t^2\,.
\end{align}

A sawtooth pattern appears because both sums compete in the time derivative of the eikonal phase. First, note that the bounds of the sums involve floors such that they are mostly constant except for discontinuities when a new image must be included. Therefore, except at these cusps in $\delta$, a time derivative only acts on the summands. This means that the time derivative of the sum $\mathbb{I}$ is reduced to $O(\Delta t^2)$, at which order it will have to compete with the time derivative of $\mathbb{II}$. Indeed,
\begin{align}
    \frac{d \delta}{d \Delta t} \sim
    \frac{4  G_N m_V m_W }{ \varepsilon^2}\left(
    1
    +4 \pi r_+ \frac{ e^{2 r_+ \big[ \left(\Delta t -\Delta \varphi - \frac{\log4 r_+ \Delta t }{r_+}\right)  ~\mathrm{mod}~2\pi  \big]} }{ e^{4 \pi r_+}-1 }
    \right)\Delta t^2
    +O(\Delta t)\,,
    \label{eq: derivative delta}
\end{align}
where the two terms in parentheses come from the two sums, respectively.
The relative size of the two terms changes depending on whether we look right before or after the cusp. These correspond to whether $\big[ \left(\Delta t -\Delta \varphi - \frac{\log4 r_+ \Delta t }{r_+}\right)  ~\mathrm{mod}~2\pi R \big]$ is just below $2\pi$ or just above $0$, respectively,
\begin{align}
    \frac{d \delta}{d \Delta t} =
    \frac{4  G_N m_V m_W }{ \varepsilon^2}
    \begin{cases}
    \left(
    1
    + \frac{ 4 \pi r_+ }{ 1-e^{-4 \pi r_+} }
    \right)\Delta t^2
    +O(\Delta t)\,,
    &\quad \big[ \left(\Delta t -\Delta \varphi - \frac{\log4 r_+ \Delta t }{r_+}\right)  ~\mathrm{mod}~2\pi  \big] \lesssim 2\pi\,, \\
    \left(
    1
    + \frac{ 4  \pi r_+ }{ e^{4 \pi r_+}-1 }
    \right)\Delta t^2
    +O(\Delta t)\,,
    &\quad \big[ \left(\Delta t -\Delta \varphi - \frac{\log4 r_+ \Delta t }{r_+}\right)  ~\mathrm{mod}~2\pi  \big] \gtrsim 0\,.
    \end{cases}
    \nonumber
\end{align}
Right before the cusp, the contribution to the derivative from $\mathbb{II}$ is biggest since $\frac{ 4 \pi r_+ }{ 1-e^{-4 \pi r_+} }>1 $ for $r_+>0$. Conversely, right after the cusp, the contribution to the derivative from $\mathbb{I}$ is biggest since $\frac{ 4 \pi r_+ }{e^{4 \pi r_+}-1 }<1 $ for $r_+>0$. At $r_+=0$ the two terms become equal, as we return to the vacuum answer. The sawtooth is sharpest when $r_+$ is large and there is a sharp hierarchy between the two contributions.

\begin{figure}[h]
\includegraphics[width=9cm]{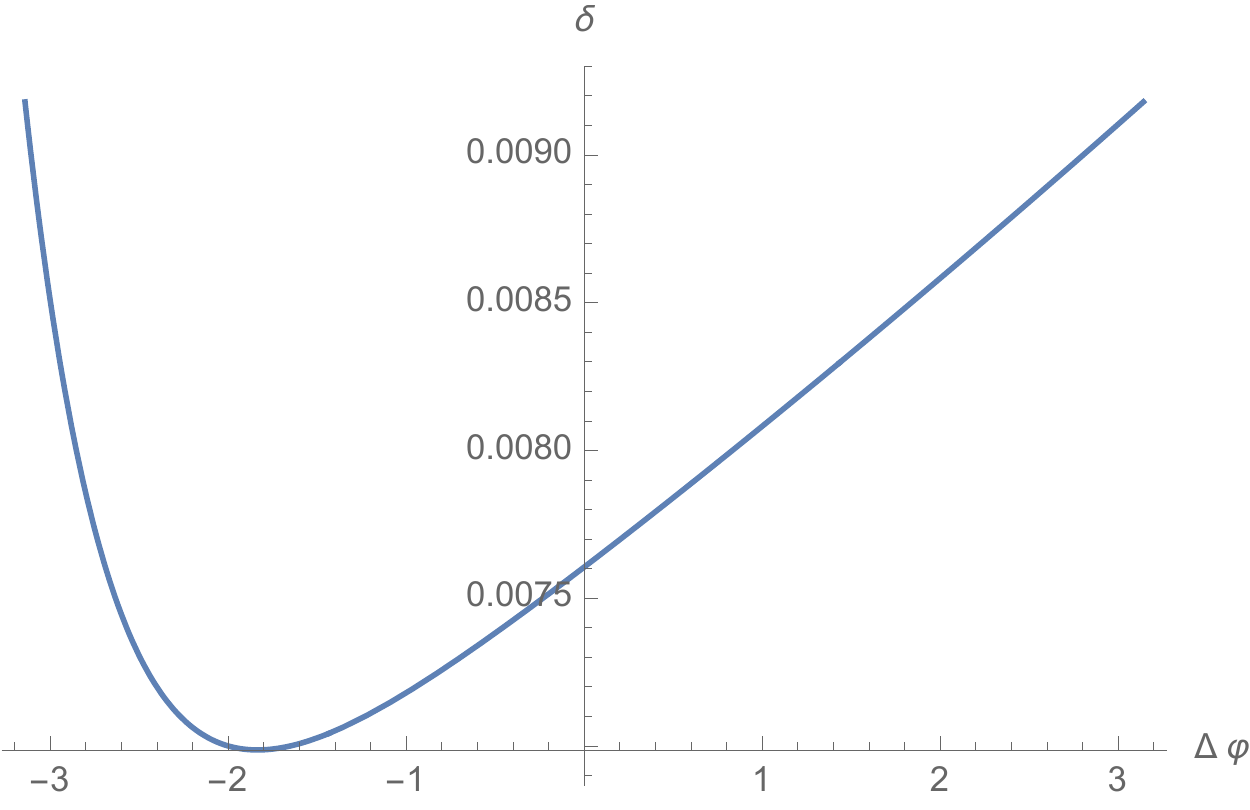}
\centering
\caption{The spatial dependence of the eikonal phase in extremal BTZ at $r_+=1$, for a fixed time separation, $\Delta t \gg 1$, corresponding to the horizontal dashed line in figure \ref{fig:contourplots}. The time separation has been chosen such that the cusp would be on the edges $\Delta \varphi = \pm \pi$. The cross-over between the exponential and power-law behavior is also visible in this figure, with the location of the trough given by \eqref{eq: cross-over}. The $\varphi$-dependence of the eikonal phase is encoded in the subleading $O(t^2)$ terms of \eqref{eqn:sum_answer}. Up to a redefinition $\phi \rightarrow \Delta t -\Delta \varphi - \frac{\log4 r_+ \Delta t }{r_+}$, this precisely matches the shape of the shockwave in the extremal limit in \cite{Mezei:2019dfv}, after substracting the divergent piece (although the latter shockwave is a priori only valid for $t \gg \beta \rightarrow \infty$).}
\label{fig: shock wave}
\end{figure}

The cross-over between the two behaviours, seen in figure \ref{fig: shock wave}, occurs when the two terms in \eqref{eq: derivative delta} are equal, i.e.
\begin{align}
     \left[ \left(\Delta t -\Delta \varphi - \frac{\log4 r_+ \Delta t }{r_+}\right)  ~\mathrm{mod}~2\pi  \right] = \pi + \frac{1}{2 r_+} \log\left(\frac{\sinh(2\pi r_+)}{2\pi r_+}\right)\,,
     \label{eq: cross-over}
     \end{align}
such that the part of the sawtooth with exponential growth decreases as $r_+$ increases. This is compatible with the fact that the rate of exponential growth is $2r_+$ and so when $r_+$ is increased the same amount of catch-up growth to interpolate between the local quadratic and overall cubic growth can be achieved in a shorter time.

\paragraph{The OTOC in non-maximally rotating BTZ.}
We would also like to comment on a connection with previous work which has studied the OTOC in rotating ensembles \cite{Jahnke:2019gxr, Mezei:2019dfv}. In those works the OTOC was computed by multiplying the center-of-mass energy with a shock wave profile computed by assuming that the scattering occurs on the horizon. In the finite temperature case, this is a good approximation in the regime $\Delta t \gg \beta$. However, in the extremal limit, this shock wave profile (denoted $f(\phi)$ in \cite{Jahnke:2019gxr} and $h(\phi)$ in \cite{Mezei:2019dfv}) diverges. This can be seen by looking at equation (5.17) in \cite{Mezei:2019dfv} for example. This is why in our approach, it was important that we did not approximate the shock wave by putting it on the horizon in the description of the gravitational scattering. Nonetheless, the shock wave profile from \cite{Mezei:2019dfv} can be regulated by subtracting a term that diverges in the near extremal limit, yet is constant in $\Delta t$ and $\Delta \varphi$.\footnote{Using the conventions of (5.17) in \cite{Mezei:2019dfv}, in the near extremal limit,
\begin{equation*}
 h(\phi) \simeq \frac{1}{2 \pi (r_+ - r_-)} + \frac12 \left( 1- \frac{(\phi ~\mathrm{mod}~2\pi)}{\pi }  +\frac{2 e^{2 r_+ (\phi ~\mathrm{mod}~2\pi)}}{e^{4 \pi r_+} -1} \right) + O\left(r_+ - r_-\right)\,.
\end{equation*}
Notice that this matches the $O(\Delta t^2)$ term in parentheses in \eqref{eqn:sum_answer} up to the replacement $\phi \rightarrow \Delta t - \Delta\varphi - \frac{\log(4 r_+ \Delta t)}{r_+}$. The constant factors multiplying the parentheses can be understood by comparing the relevant conventions and using the expression \eqref{eqn:com_ext_BTZ} for the center-of-mass energy in extremal BTZ. The variable $\phi$ in that work corresponded to a co-rotating coordinate. In the extremal limit, that co-rotating coordinate becomes null. Similarly, the expression $\Delta t - \Delta\varphi - \frac{\log(4 r_+ \Delta t)}{r_+}$ appearing in our result can be thought of as a type of co-rotating coordinate slightly regulated so that it does not become null in the extremal limit. This is also related to the retarded coordinates defined in \eqref{eq:retarded_coordinates}, where an additional logarithmic term was required compared to the similar coordinates in the non-extremal case. This term was required to ensure that a radially infalling null geodesic stays at constant $\phi$.
Alternatively, it can be understood as the term  required for the infalling coordinates to be co-rotating at the horizon rather than at the boundary. Notice that the $d\varphi dt$ term is subleading at the conformal boundary in $(t,r,\varphi)$ coordinates, \eqref{static-extremal-metric}, whereas the $d\phi dv$ vanishes at the horizon in $(v,r,\phi)$ coordinates, \eqref{eqn:infalling-coords}.
}
The regulated shock wave then corresponds to the terms of $O(\Delta t^2)$ in parentheses in \eqref{eqn:sum_answer}. This regulated shock wave does not help with computing the leading growth of $\Delta t^3$ in \eqref{eqn:sum_answer}, which arose from the sum of an interaction of strength $O(\Delta t^2)$ over $\Delta t$ images. The number of images is controlled by how deep the interaction happens in the bulk and taking the approximation that the scattering happens on the horizon would correspond to including an infinite number of these images. Since in the extremal case the shock wave has a power-law tail, this cutoff regulating the sum over images is important to track. On the other hand, the sawtooth pattern on top of this overall growth, contained in the $O(\Delta t^2)$ term, does correspond to considering the regulated extremal limit of the shock wave computed on the horizon in the non-extremal geometry times the center-of-mass energy of two colliding geodesics in extremal BTZ. The spatial dependence of the eikonal phase, which coincides with the dependence of the shockwave, is displayed in figure~\ref{fig: shock wave}.

\section{Microstate geometries}\label{section:microstate}
\label{sec:microstate}
We have seen above that the OTOC can be computed within a WKB approximation by studying the exchange of a gravitational shock wave between two boundary anchored geodesics. The strength of the interaction was controlled by the energy of the interacting geodesics in the center-of-mass frame. The form of the gravitational shock wave also played an important role in controlling the sum over images that appears due to the periodic spatial direction on the boundary.

In this section, we will consider a family of three-charge microstate geometries constructed in \cite{Bena:2016ypk,Bena:2017xbt}. These are 10-dimensional IIB supergravity solutions reduced to 6 dimensions on a $T^4$ dual to BPS states of $\mathcal{N}=4$ SYM. These geometries have the form of a 3-sphere fibred over an extremal BTZ black hole. At large radial coordinate of the BTZ base, they approach an asymptotic $AdS_3\times S^3$. At intermediate radii they have a throat region which approximates the  $AdS_2\times S^1\times S^3$ characteristic of the near horizon region of extremal BTZ. However at small radius, the throat region ends in
a smooth cap at a finite proper distance.

In section~\ref{superstrata}, we introduce the metric of these geometries. In section~\ref{section:blackholelimit}, we discuss the black hole limit of the superstrata and how the quantities derived in this section can be compared to the computations in extremal BTZ black holes. In section~\ref{section:superstratageodesics}, we study null geodesics in the superstrata, which will be required to compute the OTOC within the WKB approximation. In section~\ref{sec:OTOC_superstrata}, we use these results to discuss how the presence of the cap potentially modifies the behaviour of the OTOC in different regimes and identify a time scale where our geodesic approximation breaks down due to tidal forces. We find that these effects only become relevant after the time scale associated with scrambling when the commutator squared becomes O(1).

\subsection{(1,0,n) superstrata}
\label{sec:superstrata}
 The superstrata are solutions of six-dimensional supergravity, with
  metric given by \cite{Bena:2019azk}
\begin{align} \label{superstrata}
    \extd s_6^2=\sqrt{Q_1 Q_5}\Lambda &\Bigg[ \frac{\extd \rho^2}{\rho^2+a^2}-\frac{F_1(\rho)}{a^2(2a^2+b^2)^2 F_2(\rho)}\left(d\st-d\varphi+\frac{a^2(a^4+(2a^2+b^2)\rho^2)}{F_1(\rho)}(d\st+d\varphi) \right)^2 \nonumber \\
    &+\frac{a^2 \rho^2(\rho^2+a^2)}{F_1(\rho) }(d\st+d\varphi)^2+\extd\theta^2+\frac{1}{\Lambda^2}\sin^2\theta \left( \extd \phi_1-\frac{2 a^2}{(2a^2+b^2)}dt\right)^2 \nonumber \\
    &+\frac{F_2(\rho)}{\Lambda^2}\cos^2\theta \left( \extd\phi_2 - \frac{1}{(2a
   ^2+b^2)F_2(\rho)}\left[ -2 a^2 d\st+b^2 F_0(\rho)(d\st-d\varphi)\right] \right)^2 \Bigg],
\end{align}
where $\varphi$ goes around an $S^1$ with periodicity $\varphi\sim \varphi+2\pi$.
We have made the coordinate redefinition $t\rightarrow R_y t$ and $y \rightarrow R_y \varphi$ compared to the notation in \cite{Bena:2019azk}. This eliminates the parameter $R_y$ from their solution in order to be consistent with the conventions used in the previous section.
The following functions enter the above metric:
\begin{subequations}
\begin{align}
    F_0(\rho) &= 1-\frac{\rho^{2n}}{(\rho^2+a^2)^n} \, , \\
    F_1(\rho) &= a^6-b^2(2a^2+b^2)\rho^2 F_0(\rho) \,, \\
    F_2(\rho) &= 1-\frac{a^2b^2}{2a^2+b^2}\frac{\rho^{2n}}{(\rho^2+a^2)^{n+1}} \, , \\
    \Lambda &= \sqrt{1-\frac{a^2b^2}{2a^2+b^2}\frac{\rho^{2n}}{(\rho^2+a^2)^{n+1}}\sin^2\theta} \, .
\end{align}
\end{subequations}
In addition to the metric, there are various other supergravity fields turned on which we will not need here.

The superstrata solution is specified by the 3 parameters $(a, b, n)$. There are a number of conserved charges in supergravity which have non-zero values in this solution. The D-brane charges $Q_1$ and $Q_5$ only appear in the metric in the combination $Q_1 Q_5$, which is fixed in terms of the parameters of the superstrata solution by the regularity condition
\begin{align}
\label{eqn:reg}
    Q_1 Q_5 &= \left(a^2+\frac12 b^2\right) \, .
\end{align}
This solution also has momentum charge along the $S^1$ parametrized by $\varphi$ and angular momentum $J_L=J_R=J$ in planes that straddle both the BTZ and $S^3$ parts of the fibration. These charges are given by
\begin{gather}
    Q_P = \frac12 b^2 \,, \qquad
    J = \frac12 \mathcal{N} a^2 \,, \qquad
    \mathcal{N} = \frac{{\rm Vol}(T^4)}{\ell_{10}^8} \,,
\end{gather}
where ${\rm Vol}(T^4)$ is the normalised volume of the $T^4$ as defined in \cite{Bena:2019azk} and $\ell_{10}$ is the 10-dimensional Planck length. An effective 6-dimensional Planck length can be introduced
\begin{align}
    \ell_6^4 \equiv \frac{\ell_{10}^8}{{\rm Vol}(T^4)}\,,
\end{align}
in terms of which
\begin{align}
    J = \frac{a^2}{2 \ell_6^4} \,.
\end{align}

These supergravity solutions have well understood CFT duals which are described in \cite{Bena:2016ypk,Bena:2019azk}. Their central charge is controlled by two quantised numbers $N_1$ and  $N_5$ corresponding to the number of units of D-brane charge in the supergravity solution. We will not discuss these CFTs in detail here except to note that the supergravity charges $Q_{1,5}$ are related to the central charge of the CFT by
\begin{align}
   c =  6 N_1 N_5 =  6 \frac{Q_1 Q_5}{\ell_6^4 }\,.
\end{align}
Note that we can also relate the central charge to the three-dimensional Newton constant through the Brown-Henneaux formula \cite{Brown:1986nw},
\begin{align}
\label{eqn:Brown-Henneaux}
    c=\frac{3}{2 G_N}\,. \end{align}

\subsection{Black hole limit of the superstrata} \label{section:blackholelimit}
 We  now study the $a \rightarrow 0$ limit of the superstrata, where it approaches the extremal BTZ black hole. This allows the parametrization of the superstrata used so far to be related to $r_+$ parametrizing the extremal BTZ black hole.

In the limit $a\rightarrow 0$, the metric \eqref{superstrata} becomes
\begin{align} \label{eqn:a0limit}
    \extd s^2 &= \sqrt{Q_1 Q_5} \left[  \frac{\extd \rho^2}{\rho^2} - \frac{2 \rho^2}{b^2 }(d\st^2-d\varphi^2) + n (d\st-d\varphi)^2  + \extd\theta^2+\sin^2\theta\extd\phi_1^2+\cos^2\theta\extd\phi_2^2 \right] \, .
\end{align}
To make contact with the extremal BTZ black hole in the form
\eqref{static-extremal-metric},
one can use the regularity condition \eqref{eqn:reg} and the fact that $a \ll b$,
to identify
\begin{align}
 \rho^2 = \frac{b^2}{2} (r^2 - n) \, , \quad
    n = r_+^2 \, ,  \qquad
    Q_1 Q_5 =   \ell_{AdS}^4 \simeq \frac12 b^2\,.
\end{align}
In other words, $b$ controls the overall scale of the geometry through $\ell_{AdS}$ and $n$ controls the right-moving temperature \eqref{eq:temp} of the approximate extremal BTZ geometry,
\begin{align}
    T_R = \frac{\sqrt{n}}{\pi} \,.
\end{align}

At the conformal boundary, $r \rightarrow \infty$, $t$ is equal to the dimensionless time on the boundary measured in units of the radius of the boundary circle (taken to be unity), while $\varphi$ is an angular coordinate on that same circle.
To be consistent with the previous section, we should set $\ell_{AdS} = 1$ which means that $b = \sqrt{2}$, however since the existing literature on these geometries keeps $b$ in expressions we will do so as well.

The remaining parameter, $a$, controls the deviations from this extremal BTZ. More concretely, let us investigate the region where the superstrata closely approximates the throat of extremal BTZ.
This throat sets in for
\begin{align}
    \rho \ll
        \frac{\sqrt{n} b}{\sqrt{2}}  \equiv \rho_{throat}\,,
\end{align}
which is where the radius of the $S^1$ is approximately constant. In BTZ, this translates to $r^2 - r_+^2 \ll  r_+^2$.
It provides a good approximation for the geometry as long as $\rho \gg \sqrt{n} a$, after which the radius of the throat starts to shrink again until it pinches off at the tip of this cap region as described in detail in \cite{Bena:2019azk}. Therefore the relevant length scale associated with the start of the cap region is
\begin{align}
    \rho_{cap} \equiv \sqrt{n} a \,.
\end{align}
The proper length along the radial direction of the throat region, depicted in figure \ref{fig:microstate_cap}, is
\begin{align}
    (Q_1 Q_5)^{\frac14} \log \left( \frac{\rho_{throat}}{\rho_{cap}} \right)\,.
\end{align}
The ratio controlling the depth of the throat can also be expressed as
\begin{align}
    \frac{\rho_{throat}^2}{\rho_{cap}^2} = \frac{b^2}{2 a^2} =  \frac{
        Q_1 Q_5}{2 \ell_6^4 J} = \frac{N_1 N_5}{2 J}\,.
\end{align}

For a fixed value of the central charge, or equivalently $N_1N_5$, the longest throats are obtained by setting $J$ to be as small as possible. Since it is quantised, this is $J=\frac12$ \cite{Bena:2019azk}. This corresponds to the regime of superstrata parameter space which is closest to the extremal BTZ ensemble, since it has the minimal extra angular momentum in the extra dimensions. In the following, we set $J=\frac12$.

With this top-down understanding of how an extremal BTZ appears as a limit of these microstate geometries, the scrambling time computed in \eqref{eqn:scrambling_time_3d} can be expressed in terms of the parameters of the dual CFT state,
\begin{align}
    \Delta t_s = \left(
    \frac{ 3 \varepsilon^2 N_1 N_5 }{8 h_V h_W}
    \right)^{\frac13}\,.
\end{align}

Note that in this section we have worked in the regime $a\ll b$, where the physical identification of a long extremal BTZ throat makes sense. In extending the definition of these parameters away from this regime, it may be natural to include additional terms that are subleading in the $a\ll b$ limit.

\subsection{Geodesics of the superstrata} \label{section:superstratageodesics}
The geodesic method for computing the OTOC, outlined in section~\ref{section:geodesic} and applied in section~\ref{section:BTZ} to extremal BTZ, involves an ingoing and an outgoing null geodesic that are both anchored on the boundary and interact in the bulk of the geometry. In this section, we will analyze such geodesics in the microstate geometries described by \eqref{superstrata}. 
Geodesics in the superstrata geometries were studied in \cite{Bena:2018mpb,Bianchi:2020des}.
The present analysis is based on the results of \cite{Bena:2018mpb}, which analyzed timelike geodesics dropped into the throat region of these geometries.
As was done there, we will also restrict ourselves to geodesics with $\theta=\tfrac{\pi}{2}$, which is a fixed point of the $\theta\rightarrow\pi-\theta$ symmetry.

Working in $(\st,\varphi,\rho,\theta,\phi_1,\phi_2)$ coordinates, the Killing vectors associated to the isometries of the metric \eqref{superstrata} are given by $\partial_{\st}$, $\partial_\varphi$, $\partial_{\phi_1}$ and $\partial_{\phi_2}$, with associated conserved momenta
\begin{align} \label{momentasuperstrata}
    E=-(\partial_{\st})^\mu k_\mu \, , \quad
    P_\varphi=(\partial_\varphi)^\mu k_\mu \, , \quad
    L_1=(\partial_{\phi_1})^\mu k_\mu \, , \quad
    L_2=(\partial_{\phi_2})^\mu k_\mu \, ,
\end{align}
where $x^\mu(\tau)$ is a parametrization of the geodesic in question and $k^\mu=d x^\mu/ d\tau$.
We set $P_\varphi=0$ as was done in the previous section when studying the extremal BTZ black hole. We also set $L_1=0$ and $L_2=0$ so that the geodesic does not have extra angular momentum in the $S^3$.

The equations~\eqref{momentasuperstrata} can be solved to express components of the velocity in terms of the energy~$E$,
\begin{subequations}
\label{eq:k-micro}
\begin{align}
     \frac{\extd \st}{\extd\tau} &= \frac{(2a^2+b^2)((2a^2+b^2)(\rho^2+a^2)^n-b^2\rho^{2n})}{2\sqrt{2}a^2(\rho^2+a^2)^{n+1}\sqrt{(2a^2+b^2)-a^2b^2\rho^{2n}(\rho^2+a^2)^{-(n+1)}}}E \label{dtdtau} \,,\\
     \frac{\extd \varphi}{\extd\tau} &= \frac{(2a^2+b^2)b^2(\rho^2+a^2)^{-\frac12 (n+1)}((\rho^2+a^2)^n-\rho^{2n})}{2\sqrt{2}\sqrt{(2a^2+b^2)(\rho^2+a^2)^{n+1}-a^2 b^2 \rho^{2n}}}E \,,\\
     \frac{\extd \phi_1}{\extd\tau} &=  \frac{(2a^2+b^2)(\rho^2+a^2)^n-b^2\rho^{2n}}{\sqrt{2(\rho^2+a^2)^{n+1}}\sqrt{(2a^2+b^2)(\rho^2+a^2)^{n+1}-a^2 b^2 \rho^{2n}}} E \,,
\end{align}
\end{subequations}
and $\extd \theta/\extd\tau = \extd\phi_2/\extd\tau=0$. The condition that the geodesic be null, $k^2=0$, can be solved for the remaining component of the velocity
\begin{align} \label{drdtausq}
    \left(\frac{\extd \rho}{\extd\tau}\right)^2 &= \frac{2a^2+b^2}{2a^2}\frac{(\rho^2+a^2)((2a^2+b^2)(\rho^2+a^2)^{n}-b^2\rho^{2n})}{(2a^2+b^2)(\rho^2+a^2)^{n+1} - a^2 b^2 \rho^{2n}} E^2\,.
\end{align}

In order to find the trajectory of the geodesics we simply need to integrate the ratio of the velocities \eqref{dtdtau} and \eqref{drdtausq}
\begin{align} \label{dtdrsols}
    \st(\rho) - \st(\infty) = \pm \int_\rho^\infty d\rho \frac{\sqrt{2a^2+b^2}\sqrt{(2a^2+b^2)(\rho^2+a^2)^n-b^2 \rho^{2n}}}{2a(\rho^2+a^2)^{\frac{n}{2}+1}} \,.
\end{align}
This expression gives the coordinate time at which a null geodesic emitted from the boundary will probe a given radius of the geometry. Now consider two geodesics, one which leaves the boundary at $\st_{in}$ and one that is absorbed at the boundary at $\st_{out}$.
These two geodesics will rotate around the compact $\varphi$ and $\phi_1$ directions, but they will most strongly interact in the bulk when the ingoing and outgoing geodesics meet at the same radius, $\rho_*$.

A full computation of their interaction would require computing the dependence on the compact directions of the gravitational shock wave produced by geodesics in the superstrata geometry. However, here we will simply take the first steps towards understanding the difference between the superstrata and the extremal BTZ black hole. We will therefore focus on the motion in the $\rho$-$\st$ directions in order to better understand the effect of the cap in the superstrata geometry without the additional complication of the effects of these compact directions.

The radius where the two geodesics meet, $\rho_*$, is related to the difference of their insertion times at the boundary by
\begin{align}
\label{eqn:Delta_t}
    \Delta \st (\rho_*) \equiv \st_{out}-\st_{in} = 2 \int_{\rho_*}^\infty d\rho \frac{\sqrt{2a^2+b^2}\sqrt{(2a^2+b^2)(\rho^2+a^2)^n-b^2 \rho^{2n}}}{2a(\rho^2+a^2)^{\frac{n}{2}+1}} \,.
\end{align}
We will now proceed to study this expression in various regimes. 

\paragraph{BTZ region.} First, let us consider the region well outside the cap, where the geometry is well approximated by extremal BTZ $\times S^3$, for $\rho_*,b \gg a$,
    \begin{align} \label{eq:Deltatouterapprox}
        \Delta \st &\simeq 2 \int_{\rho_*}^\infty \frac{\sqrt{2\rho^2+nb^2}b }{2 \rho^3} \extd \rho= \frac{b }{2 \rho_*^2} \sqrt{2\rho_*^2+nb^2} + \frac{1}{\sqrt{n}} \arcsinh{\frac{\sqrt{n}b}{\sqrt{2}\rho_*}}\,.
    \end{align}
    The OTOC will start probing the throat when $\rho_* = \rho_{throat}$, corresponding to a time separation on the boundary of
    \begin{align}
        \Delta \st_{throat} \simeq \left( \sqrt{2}+\arcsinh{1} \right)\frac{1}{\sqrt{n}}  \implies \Delta t_{throat} \propto T_R^{-1} \, .
    \end{align}
    This timescale is controlled by the right-moving temperature of the extremal BTZ black hole approximated by this geometry. As expected, this time scale only depends on BTZ parameters and does not involve details of the cap region.
    
In this regime the motion of the geodesic along the BTZ factor of the geometry closely follows the trajectory of a geodesic in extremal BTZ, that is $\big( t(\rho),\varphi(\rho) \big)_{superstrata} \simeq \big( t(\rho),\varphi(\rho) \big)_{BTZ}$. In this asymptotic region, the geometry is approximately a product of BTZ and a homogeneous $S^3$. The geodesics we consider do not have any angular momentum along the $S^3$, such that they sit at a constant value of the sphere coordinates $(\theta,\phi_1,\phi_2)$.

To understand where the corrections to the geometry first start to significantly affect the trajectory, the ratio of the first subleading term to this leading term must be examined. In the regime where $a \ll \rho_* \ll b$, by expanding $t(\rho)$ in \eqref{eqn:Delta_t} to subleading order, we find that this ratio is
\begin{align}
   -\frac{(n+5) a^2}{8 \rho^2}     \,.
\end{align}
By using a similar approach to compute $\varphi(\rho)$, we find the same ratio.
The perturbative expansion used in \eqref{eq:Deltatouterapprox} breaks down when this ratio is $O(1)$ at $\rho \sim \sqrt{n+5}\, a \sim \rho_{cap} $. 

\paragraph{Cap region.}In order to access the cap region, we can instead expand \eqref{eqn:Delta_t} in the regime where $\rho_*,a \ll b$. To do so, we must split the integral into two parts somewhere in the overlap of the applicability of the two approximations, at a radius $\rho_{split}$ obeying both  $\rho_{split},b \gg a$ and $\rho_{split},a \ll b$, i.e. $a \ll \rho_{split} \ll b $, 
\begin{align}
\label{eqn: approx dt}
    \Delta \st &\simeq 2 \int_{\rho_{split}}^\infty \frac{\sqrt{2\rho^2+nb^2}b }{2 \rho^3} \extd \rho + 2 \int_{\rho_*}^{\rho_{split}} \frac{\sqrt{(\rho^2+a^2)^n-\rho^{2n}}b^2 }{2a(\rho^2+a^2)^{\frac{n}{2}+1}} \extd \rho \simeq  \frac{ b^2 }{a^2} \chi_n\left(\frac{\rho_*}{a}\right)
    \end{align}
  where only the leading term in $\frac{b}{a} \gg 1$ has been kept and the following dimensionless integral was introduced,
    \begin{align}
    \label{eqn:chi_def}
        \chi_n(x) &\equiv \int_{x}^{\infty} \frac{\sqrt{(\xi^2+1)^n-\xi^{2n}}}{ (\xi^2+1)^{\frac{n}{2}+1}}\extd\xi  \,.
    \end{align}

    The minimal separation between the insertions on the boundary so that the OTOC can directly probe the cap occurs when $\rho_* = \rho_{cap}$, so that
    \begin{align}
        \Delta \st_{cap} \simeq \frac{b^2 }{a^2} \chi_n(\sqrt{n})\,.
    \end{align}
    At large $x$,
   $  \chi_n(x)\simeq \frac{\sqrt{n}}{2x^2} $,
    so that at large $n$, $\chi_n(\sqrt{n}) \simeq \frac{1}{2\sqrt{n}}$. In fact, numerical investigation shows that for any $n \geq 1$,
    \begin{align}
        \frac14 < \sqrt{n}\, \chi_n (\sqrt{n}) < \frac12\,.
    \end{align}
    Therefore the time scale to reach the cap region is given by
    \begin{align}
\label{eqn:t_cap}
      \Delta t_{ cap}  \propto \frac{1}{\sqrt{n}} \frac{b^2}{a^2}
                \propto \frac{N_1N_5}{T_R}\,,
                   \end{align}
reproducing the result quoted in the introduction in \eqref{eqn:t_cap-intro}.

\subsection{OTOC in the superstrata}
\label{sec:OTOC_superstrata}
Now that we have understood the trajectory of the relevant geodesics, we can see how this affects the computation of the OTOC in the superstrata geometry. First of all, we should be clear about the OTOC we wish to compute. In the asymptotic region, reducing on the compact $S^3 \times T^4$ of the 10-dimensional geometry leads to a Kaluza-Klein tower of fields from the 3-dimensional BTZ perspective. Each of the 3-dimensional fields in this tower corresponds to an operator in the dual 2-dimensional CFT. We will restrict ourselves to considering OTOCs of operators dual to the lowest field in this tower of Kaluza-Klein modes, which are completely delocalised along these extra compact dimensions. 
More precisely, this means that in the bulk-to-boundary propagators used to compute the OTOC, such as in equation \eqref{eq:WKB-propagator}, the boundary insertion point is integrated over $S^3 \times T^4$. In the part of the geometry well approximated by extremal $\mathrm{BTZ} \times S^3 \times T^4$, this leads to fields which are completely homogeneous in the compact dimensions and which source a homogeneous shock wave. This is compatible with the fact that the geodesics in this region with no angular momentum along the $S^3$ follow BTZ geodesics as discussed in the previous section, so that the WKB approximation applied to the 3-dimensional reduction is consistent with the 6-dimensional picture. With this set up in mind, there are three major effects which will cause the computation of the OTOC in the superstrata to differ from the computation in extremal BTZ and we will discuss each one in turn.

First, the center-of-mass energy of the interacting geodesics will be modified once the bottom of the throat is reached. In BTZ, the center-of-mass energy continues to grow in an unbounded fashion as the interaction gets closer and closer to the horizon. In the superstrata, this growth is cut off by the depth of the throat once the interaction moves into the cap region. Below, we will compute the center-of-mass energy of two colliding geodesics as they fall down the throat of the superstrata and see how the growth in this quantity saturates. This will occur at the time scale set by $\Delta t_{cap}$. This time scale can be compared to the scrambling time, to determine whether the effect of the cap will be felt before scrambling,
\begin{align}
    \frac{\Delta t_{cap}}{\Delta t_s} = \left( \frac{8 h_V h_W (N_1 N_5)^2}{3 \varepsilon T_R^3} \right)^{\frac13} \,.
\end{align}
We conclude that the OTOC will have exited the slow scrambling regime before the interaction reaches the cap, unless the right-moving temperature is very large,
\begin{align}
    T_R \gtrsim \left( \frac{h_V h_W (N_1 N_5)^2}{ \varepsilon}\right)^{\frac13} \,.
\end{align}

Second, the shock wave will be modified by the presence of the cap.
The geometry away from the cap is well approximated by a geometry with an extremal BTZ factor. Within this part of the geometry, all the fields remain homogeneous in the extra compact dimensions. Therefore the gravitational shock wave does not depend on these extra dimensions and the BTZ result can be used. In particular, the average time growth is cubic since the number of images contributing to the gravitational interaction grows linearly while each individual contribution is proportional to the center-of-mass energy $s \sim \Delta t^2$. Once the cap region is reached, the fibration of the $S^3$ over BTZ becomes non-trivial and the shock wave must be computed using the full 6-dimensional geometry.\footnote{Note that the OTOC of an operator localised in the $S^3$ would correspond to a computation using a single geodesic in the 6-dimensional geometry with no smearing and the resulting gravitational shock wave would propagate in the $S^3$. From the perspective of the boundary CFT, localising the insertion in the $S^3$ requires a sum over a tower of primary operators so this is not the most natural OTOC to consider.}

In the following, we will argue more carefully that the deformation of the geometry in the cap region does not affect the gravitational shock wave relevant to the OTOC at times before the interaction reaches the cap region. Consider the time slice where the first operator is inserted at $\st_{in}$. On this slice we can find a solution to Einstein's equations that is compatible with this source and that is only supported near the boundary. As we evolve time forward, the shock wave will not probe deeper into the geometry than the null geodesic that sources it (the support of the shock wave is depicted in figure~\ref{fig: shock wave 3D} in appendix~\ref{appendix:shock wave} for non-compact $\varphi$). We can stop evolving time once we reach the interaction region. If this interaction occurs well outside the cap, $\rho_* \gg a$, the shock wave we needed to consider is only supported in the region where $\rho\gg a$. Since we know that this shock wave is a solution to Einstein's equations with a source at the geodesic for an extremal BTZ background and that the metric of the superstrata for $\rho \gg a$ is well approximated by extremal BTZ, then the shock wave will still be a solution in the superstrata up to subleading terms in $a/\rho$ and $a/b$. The same argument can be run in reverse for the shock wave sourced by the outgoing geodesic, by starting at the time where it reaches the boundary, $\st_{out}$ and evolving the shock wave backwards in time to the interaction region.

Once the interaction reaches the cap region, this approximation will no longer be valid. Since the number of images included in the sum was essentially controlled by boundary causality, one may not expect this to change the overall scaling with $\Delta t$. However, the detailed form of the OTOC once the interaction reaches the cap region will be affected by the precise form of the shock wave in the 6 dimensional geometry. We will not compute this precise form in this work, but we are optimistic that it may be tractable to do so due to the successes in computing two-point functions in the superstrata in \cite{Bena:2019azk} thanks to the approximately separable form of the wave equation.

Finally, \cite{Tyukov:2017uig,Bena:2018mpb,Bena:2020iyw,Martinec:2020cml} found that probes falling into the throat of the superstrata feel strong tidal forces well before reaching the cap region. In fact, these strong tidal forces  potentially invalidate the WKB approximation we have made in computing the OTOC.
In particular, we expect the expansion of the geodesic congruence to quickly grow in the presence of strong tidal forces, violating the assumption that it is much smaller than the mass of the probe required for the WKB approximation to hold, as described in more detail below.
Below, we adapt their computation to the present setup and find that the tidal stresses are in danger of invalidating the WKB approximation
at a time scale $\Delta t_{tidal}$, given in \eqref{eqn:t_tidal},
well before the center-of-mass energy of the interaction starts to saturate at $\Delta t_{cap}$.
Since we lose control of the computation beyond $\Delta t_{tidal}$, we cannot say for certain what the behaviour of the OTOC will be past this time. We expect, however, that whatever happens beyond this point, the OTOC cannot continue to grow beyond the bound identified below in \eqref{eqn:s_bound}, since the center-of-mass energy, which controls the strength of gravitational interactions, is bounded by the presence of the cap in any case.

In the above discussion, we have focused on boundary time scales. In the bulk, it is more natural to think instead of the radial coordinate at which the interaction occurs.
Although the cap deforms the near-horizon region, in the careful treatment of the OTOC needed for zero temperature states, we find that the OTOC is controlled by an interaction centered  at a small but finite distance from the horizon. The scrambling time is reached when this interaction is at a radial coordinate
\begin{align}
r_{sat} - r_+ \sim \left( \frac{G_N m_W m_V }{\varepsilon^2} \right)^{ \frac13}\,,
\end{align}
whereas the tidal forces become important at a radial coordinate
\begin{align}
r_{tidal} - r_+ \sim \sqrt{ \frac{G_N (\pi^2 T_R^2+1)  }{ \pi T_R \,\varepsilon  \min(h_V,h_W)   }} \,.
\end{align}
these are both intermediate distance scales between the Planck and AdS scales away from the horizon.

\begin{figure}[h]
\includegraphics[width=12cm]{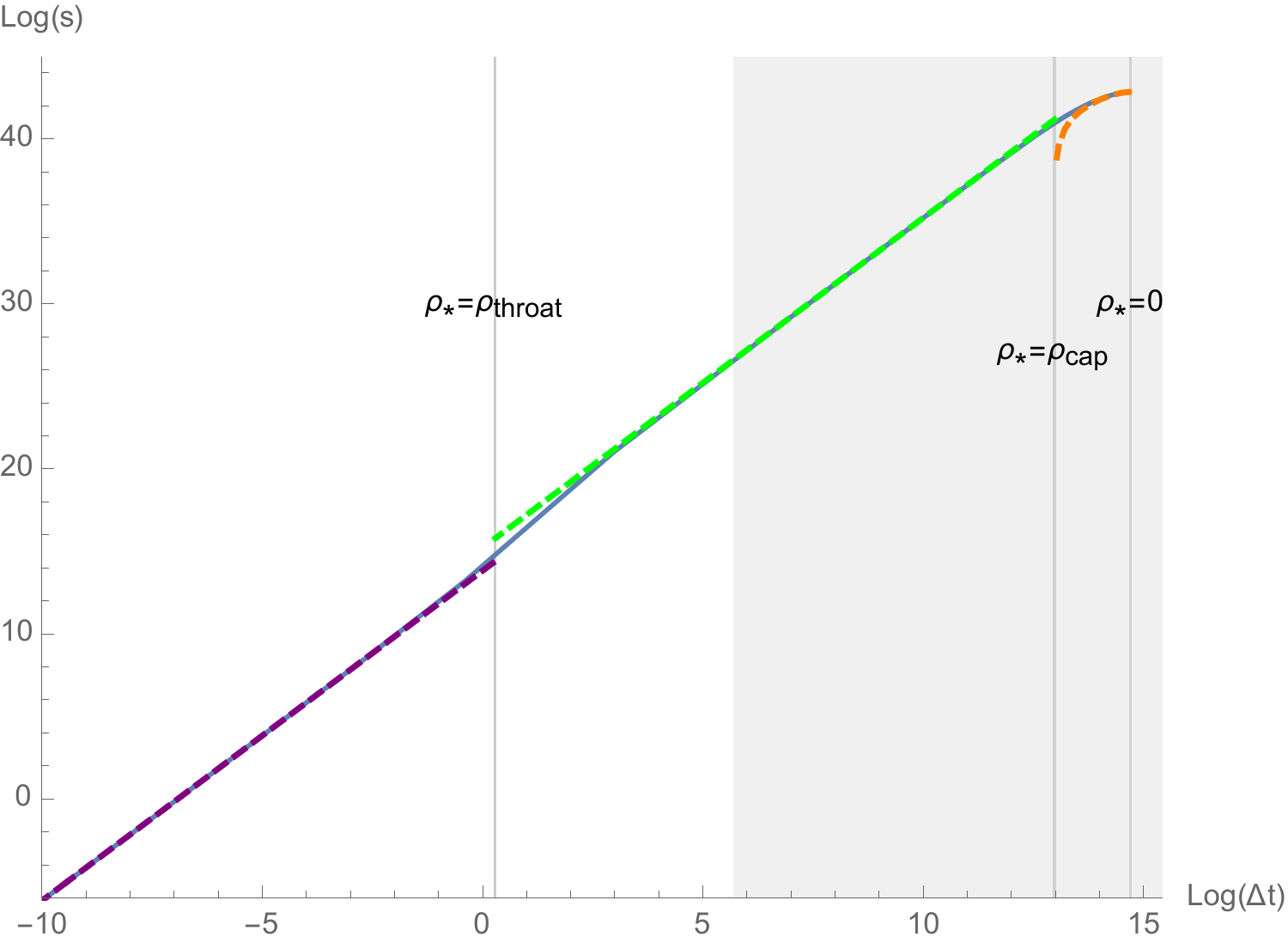}
\centering
\caption{The solid blue line is the full numeric answer for the center-of-mass energy, while the dashed lines are the analytic approximations in the respective regimes. In the grey area, the tidal forces become large. The parameters used are $n=3$, $a=10^{-3}$, $b=\sqrt{2}$, $m_V=m_W=10$ and $\varepsilon=10^{-2}$.}
\label{fig:s_superstrata}
\end{figure}

\paragraph{Center-of-mass energy.}
\label{sec:s_microstate}
Given the relationship between the velocities and the conserved quantities in \eqref{eq:k-micro} and \eqref{drdtausq}, the center-of-mass energy can be expressed in terms of the conserved quantities and the radius, $\rho_*$, where the collision occurs,
\begin{align}
    s&=         -\left(m_V k_V + m_W k_W\right)^2\big|_{\rho=\rho_*} \,,\\
    &=\frac{ b(2a^2+b^2)\left((2a^2+b^2)(\rho_*^2+a^2)^n-b^2 \rho_*^{2n}\right)}{a^2 (\rho_*^2+a^2)^{n+1}\sqrt{2a^2+b^2-a^2 b^2 \rho_*^{2n}(\rho_*^2+a^2)^{-(n+1)}}}\frac{m_V m_W} {\varepsilon^2} + O(\varepsilon^0) \, .
\end{align}
where $m_V$ and $m_W$ are the masses of the in- and outgoing geodesics respectively and we set $E_V=E_W=\sqrt{\frac{b}{\sqrt{2} }}\varepsilon^{-1}$ where we have restored the factors of $b$ in the energy computed in \eqref{eqn:energy}.
In order to make the connection with the OTOC, the location of the collision must be re-expressed in terms of the times at which the perturbations are inserted at the boundary.
This is straightforward in principle by inverting \eqref{eqn:Delta_t}, however it is not tractable to perform this inversion analytically.
The relation between $s$ and $\Delta t$ is plotted in figure \ref{fig:s_superstrata}.

We can find approximate answers by dividing the microstate geometry into three regions: an asymptotic region $\rho\gg\sqrt{n}\,b$, a throat region $\sqrt{n}\,a \ll \rho \ll \sqrt{n}\,b$ and a cap region $\rho \ll \sqrt{n}\,a$. We emphasize again that the overall time growth of the OTOC also depends on the number of images contributing to the gravitational interaction. As argued above, the BTZ result should apply in the asymptotic and throat regions, yielding an additional factor of $\Delta t$. With this in mind, below we only study the behavior of the center-of-mass energy.
\begin{itemize}
    \item In the asymptotic region, $\rho_*\gg\sqrt{n}\, b\gg\sqrt{n}\, a$, we can approximate the integral \eqref{eqn:Delta_t} by
    \begin{align}
        &\Delta \st \simeq 2\int_{\rho_*}^\infty d\rho \frac{b }{\sqrt{2}\rho^2}  = \frac{\sqrt{2} b }{\rho_*} \, .
    \end{align}
    For the center-of-mass energy this yields quadratic growth in this region:
    \begin{align}
        & s \simeq 2 \frac{b^2 }{\rho_*^2}\frac{m_V m_W} {\varepsilon^2} \simeq \frac{m_V m_W} {\varepsilon^2} \Delta \st^2
        = \frac{4\sqrt{2}}{b}\frac{h_V h_W} {\varepsilon^2}\Delta \st^2 \, ,
    \end{align}
    In the last equality we have used $2h_{V,W}=\sqrt{b/\sqrt{2}}\, m_{V,W}$ where we have again restored factors of $b$ compared to section \ref{section:BTZ}. 
    
    \item In the throat region, $\sqrt{n}\, a \ll \rho_* \ll \sqrt{n}\, b$, we get
    \begin{align}
    \label{eqn:t_of_rho_throat}
        \Delta \st &\simeq \Delta \st_{throat} +  2\int_{\rho_*}^{\sqrt{\tfrac{n}{2}}b} d\rho \frac{\sqrt{n}b^2 }{2\rho^3}
                \simeq \frac{\sqrt{n}   b^2}{2 \rho_*^2}\,,
    \end{align}
    which, for the center-of-mass energy, yields
    \begin{align}
        s & \simeq \frac{n b^4 }{\rho_*^4}\frac{m_V m_W} {\varepsilon^2}
= \frac{16\sqrt{2}}{b}\frac{h_V h_W}{ \varepsilon^2 } \Delta \st^2 \,.
    \end{align}
    This leads again to a growth in the center-of-mass energy that is quadratic in time.

\item In the deep cap region, $\rho_* \ll \sqrt{n} \, a \ll \sqrt{n} \, b$, we get
    \begin{align}
        \Delta \st
        &\simeq \Delta \st_{bottom} - 2\int_0^{\rho_*} d\rho \frac{b^2 }{2 a^3}
                \simeq  \frac{b^2}{a^2} \left( \chi_n(0) - \frac{\rho_*}{a} \right) \,,
            \end{align}
    where $\Delta \st_{bottom}$ was approximated using \eqref{eqn: approx dt}
    \begin{align}
        \Delta \st_{bottom} \equiv 2\int_0^\infty \frac{d\st}{d\rho} \simeq \chi_n(0)\frac{b^2}{a^2} \sim \frac{b^2}{a^2} \, ,
    \end{align}
    and where $\chi_n(0)$ was estimated by using
    \begin{align}
       \frac{1}{2\sqrt{n}} < \frac{\pi}{2} - \chi_n(0) < \frac{2}{3\sqrt{n}} \,,
    \end{align}
     which can be verified numerically.
    The behaviour of the center-of-mass energy in this regime is then approximated by
    \begin{subequations}
    \begin{align}
        & s \simeq
        \begin{cases}
        \frac{(2a^2-3\rho_*^2)b^4 }{2a^6}\frac{m_V m_W} {\varepsilon^2} & \text{for}\quad n=1\\
        \frac{(a^2-\rho_*^2)b^4 }{a^6}\frac{m_V m_W} {\varepsilon^2} & \text{for}\quad n>1
        \end{cases} \\
        & \simeq
        \begin{cases}
        \left(\frac{b^4 }{a^4} - \frac{3}{2}(\Delta \st_{bottom}-\Delta \st)^2 \right)\frac{m_V m_W} {\varepsilon^2} & \text{for}\quad n=1 \\
        \left(\frac{b^4}{a^4} - (\Delta \st_{bottom}-\Delta \st)^2 \right)\frac{m_V m_W} {\varepsilon^2} & \text{for}\quad n>1
        \end{cases}
    \end{align}
    \end{subequations}
\end{itemize}

At the bottom of the cap, the center-of-mass energy is given by
\begin{align}
\label{eqn:s_bound}
s_{bound} \simeq \frac{m_V m_W b^4} {\varepsilon^2 a^4} = \frac{16\sqrt{2}}{b} \frac{h_V h_W (N_1 N_5)^2}{\varepsilon^2}\,.
\end{align}
After the geodesics reach the bottom of the cap, they bounce off and start to move back up the throat. This therefore gives an upper bound on the center-of-mass energy of the interaction in the superstrata geometry.

These various approximate regimes are compared to the numeric answer in figure \ref{fig:s_superstrata}.

\paragraph{Tidal stress.}\label{subsection:Scaleoftidalstresses}
The geodesic approximation to field propagation, used in this work to compute OTOCs, has a limit of validity. Indeed, as any WKB-type approximation, it rests on the assumption that the amplitude variation of the wavefunction \eqref{eq:WKB-propagator} does not compete with that of its phase factor \eqref{eq:WKB-S},
\begin{equation}
\label{eq:WKB-validity}
|\partial_\mu A| \ll m |k_\mu A|  \,,
\end{equation}
where this condition must hold for each of the components individually, and so in particular it must hold for the component along the velocity $k^\mu$ of the geodesic,
\begin{align}
    \frac{d}{d\tau}A\equiv k^\mu \partial_\mu A\,.
\end{align}
By expanding the Klein-Gordon equation to first subleading order in $1/m$, it may be shown that the variation of the amplitude~$A$ along $k^\mu$ is related to the \textit{volume expansion} $\theta$ of a congruence of geodesics \cite{Balasubramanian:2019stt}
\begin{equation}
\frac{d}{d\tau}A=-\frac{\theta}{2} A, \qquad \theta\equiv \nabla_\mu k^\mu.
\end{equation}
Therefore, the validity of the geodesic approximation through \eqref{eq:WKB-validity} requires in particular
\begin{equation}
\label{eq:WKB-validity-2}
\left|\theta \right|\ll m.
\end{equation}

The evolution of the expansion $\theta$ is controlled by the Raychaudhuri equation
\begin{align}
    \dot \theta = -\frac12 \theta^2 - \sigma_{\mu\nu} \sigma^{\mu\nu} + \omega_{\mu\nu}\omega^{\mu\nu} - R_{\mu\nu} k^\mu k^\nu\,,
\end{align}
where $\sigma_{\mu\nu}$ and $\omega_{\mu\nu}$ are the \textit{shear} and \textit{rotation} tensors, respectively; for more details, see textbooks such as \cite{Poisson:2009pwt}.
The quantity $R_{\mu\nu} k^\mu k^\nu$ is known as the trace of the tidal tensor and quantifies the tidal forces felt by a congruence of geodesics.
In the context of superstratum microstates, it has been shown that tidal forces on infalling particles are Planckian long before they reach the cap region \cite{Tyukov:2017uig,Bena:2018mpb,Bena:2020iyw,Martinec:2020cml}. In the presence of such Planckian tidal forces, one may quickly expect a violation of the condition \eqref{eq:WKB-validity-2} and a breakdown of the validity of the geodesic approximation.

The leading term in the trace of the tidal tensor can be computed in the throat region of the geometry, where $\sqrt{n} a\ll \rho \ll \sqrt{n} b$,
\begin{align}
   - R_{\mu\nu} k^\mu k^\nu \sim -\frac{2 \sqrt{2}}{b}- \frac{n(n+1) a^2 b^2 E^2 }{\rho^6}
   =-\frac{2 \sqrt{2}}{b} -  \frac{n(n+1) a^2 b^3 }{\sqrt{2} \varepsilon^2 \rho^6}\,,
\end{align}
where all higher order terms in the small parameters $\frac{\rho}{b}$, $\frac{a}{\rho}$ and $\frac{a}{b}$ have been dropped. These two terms can compete depending on how deep into the throat we look. The first term does not depend on $a$ and so it is what we would get for an extremal BTZ black hole. For $\rho \sim \sqrt{n} b$ the second term is suppressed by $\frac{a^2}{b^2} \sim (N_1 N_5)^{-1}$ and so it is very small and cannot cause a large expansion. However, as the geodesic falls into the throat this second term grows and eventually gets bigger than the first.

Assuming that our congruence of geodesics obeys the condition \eqref{eq:WKB-validity-2} for BTZ,\footnote{We do not generically expect the condition \eqref{eq:WKB-validity-2} to be violated in the near horizon region of BTZ. Since the geometry is locally AdS, the tidal tensor is never large in BTZ (this fact was pointed out by \cite{Tyukov:2017uig}). Also, neglecting the shear and rotation in the Raychaudhuri equation allows it to be integrated in BTZ and the solutions are approximately constant in the throat region.} the danger comes from the additional term in the tidal tensor that grows as the geodesic falls down the throat. As this additional term grows, it causes $\dot\theta$ to grow, which will cause a potential violation of the condition \eqref{eq:WKB-validity-2} when
\begin{align}
   \left|
   \int \frac{n(n+1) a^2 b^3 }{\sqrt{2} \varepsilon^2 \rho^6} d\tau
   \right| \sim m \,.
\end{align}

We change variables from $\tau$ to $\rho$ in this integral by using our known expressions for the velocity of the geodesic. Expanding \eqref{drdtausq} in the throat regime,  $\sqrt{n} a\ll \rho \ll \sqrt{n} b$,
\begin{align}
   \frac{\extd \rho}{\extd\tau}
   =
   -\frac{\sqrt{n}\, b^{\frac32} }{2^{\frac34} \varepsilon \rho}\,.
\end{align}

This will cause \eqref{eq:WKB-validity-2} to be violated
at a radial coordinate,
\begin{align}
    \rho_{tidal}^4 \sim \frac{a^2 b^{\frac32} \sqrt{n} (n+1) }{ 2^{\frac74} \varepsilon m  } \,.
\end{align}
Using \eqref{eqn:t_of_rho_throat}, this translates into a boundary time separation of
\begin{align}
\label{eqn:t_tidal}
    \Delta t_{tidal} \sim \sqrt{
        \frac{ \pi T_R}{ \pi^2 T_R^2+1 }
    \min(h_V,h_W) \varepsilon  N_1 N_5
    }\,,
    \end{align}
where $\min(h_V,h_W)$ denotes the minimum of the conformal weights of the operators in the OTOC and appears because the condition $\eqref{eq:WKB-validity-2}$ must be imposed on both the in- and outgoing geodesics. This reproduces the result \eqref{eqn:t_tidal-intro} quoted in the introduction.

This time scale can be compared to the scrambling time,
\begin{align}
    \frac{\Delta t_{tidal}}{\Delta t_s}
    \sim \frac{2}{3^{\frac13}}
    \sqrt{ \frac{ \pi T_R  }{ (\pi^2 T_R^2+1 ) } } \sqrt{\min(h_V,h_W)}
    \left(\frac{h_V^2 h_W^2 N_1N_5}{\varepsilon} \right)^{\frac16}\,.
\end{align}
In this expression, $N_1N_5 \gg1$ in order to have a hierarchy between the Planck and AdS scales (semi-classical regime), $\varepsilon \ll1$ since it is the holographic regulator and $h_{V,W} \gg 1$ for the validity of the WKB approximation. All these scalings contribute to ensuring that $\Delta t_s \ll \Delta t_{tidal}$ so that we can reliably approximate the superstrata by BTZ during the slow scrambling phase of the OTOC. The only  way this condition can be violated is for very small or very large right-moving temperature. Since the superstrata solutions only exist for integer $n$, the temperature is bounded from below by $\pi T_R \geq 1$ and we only need to worry about large temperatures. In that case, $\Delta t_{tidal} \lesssim \Delta t_s$ when
\begin{align}
    T_R \gtrsim \min(h_V,h_W)
    \left(\frac{h_V^2 h_W^2 N_1N_5}{\varepsilon} \right)^{\frac13}\,.
\end{align}
As far as we are aware, there is no obstruction to considering superstrata with arbitrarily large right-moving temperature. In this large right-moving temperature regime, we expect the WKB approximation to break down, leading to deviations from the BTZ result, before the scrambling time is reached.

\section{Discussion}\label{section:discussion}

In the following, we point to some open problems and possible future directions.

\paragraph{Late-time regime and quasi-normal decay.}
We have described OTOCs in extremal geometries within the geodesic approximation scheme presented in section~\ref{section:geodesic}. This approach requires improvement if it is to accurately describe the late-time regime where quasi-normal decay may occur. It is very likely that the latter coincides with a regime where the assumption \eqref{eq:WKB-validity} does not hold anymore. This would be in line with earlier study of black hole quasi-normal modes within related WKB-type approximations \cite{Festuccia:2008zx}. We also refer the reader to \cite{Birmingham:2001pj,Bena:2019azk} for a description of quasi-normal decay of two-point functions in states dual to extremal BTZ and superstratum microstates.

\paragraph{Shock waves in microstate geometries.}
As described in section~\ref{sec:OTOC_superstrata}, we expect OTOCs in superstratum geometries to coincide with those in extremal BTZ as long as the dual particle scattering does not probe some of its distinctive features, including strong tidal forces and the existence of a cap region. This assumes that shock waves emitted by particles falling from infinity in these microstate geometries do not significantly differ from their analogue in extremal BTZ, at least far away from the throat region where strong tidal forces occur. Once the interaction probes sufficiently deep into the throat, we would need to understand the shock wave in the full 6-dimensional geometry of the cap. Two-point functions have been computed in the superstrata geometry by exploiting the almost separable form of the wave equation in \cite{Bena:2019azk}, which suggests that the computation of this shock wave may be tractable.

\paragraph{Breakdown of the geodesic approximation due to tidal forces.} The analysis performed in section~\ref{sec:OTOC_superstrata} strongly suggests that tidal forces invalidate the geodesic approximation to field propagation at the timescale $\Delta t_{tidal}$ given in \eqref{eqn:t_tidal}. A fully rigorous proof of this fact would require a more detailed study of the Raychaudhuri equation describing the congruence of geodesics used to propagate bulk fields from their boundary insertion points, including the effect of shear and rotation associated to this congruence.

\paragraph{Right-moving operators.}
We have discovered that the OTOC of scalar  operators, when evaluated in a state dual to extremal BTZ, displays a time-dependence which alternates between a quadratic growth associated to the zero left-moving temperature $T_L$ and an exponential Lyapunov growth associated to the non-zero right-moving temperature $T_R$. It is interesting to contemplate the possibility that the OTOC of purely right-moving CFT operators could display a purely exponential growth. In the gravitational description, these right-moving CFT operators correspond to high spin fields with equal spin $s$ and right conformal weight $h_R$. Since the geodesic approximation holds for large masses $m\gg 1$, high spin fields would have to be considered in order to see this purely exponential Lyapunov growth at the temperature $T_R$ within this approximation.

\paragraph{Probes of higher complexity.}
Two-point functions in the superstrata were studied in \cite{Bena:2019azk}, and were found to closely approximate the decay found in extremal BTZ until a time scale $\sqrt{N_1 N_5}$, where they start to deviate. At a time scale equal to $\Delta t_{cap} \sim \frac{N_1N_5}{T_R}$, they start to grow again and exhibit a sharp echo at a time scale $t_{echo} = N_1N_5$. Thus, these probes exhibit the effects of the difference between the superstrata and extremal BTZ geometries at similar time scales as the OTOC. One might have expected more complex probes to be more sensitive to the difference between a microstate or a statistical ensemble, but that does not seem to be the case here. This may be related to the fact that the relevant time scales are powers rather than exponentials of $N_1N_5$ so that the difference between $N_1 N_5$ and $2N_1 N_5$ is lost.

The new feature of the OTOC is the appearance of a new time scale, the scrambling time.
We expect more complex probes to take longer to scramble and so it would be interesting to compare their scrambling time to the time where they start to feel the effects of the microstate.

\paragraph{Extremal black holes in higher spacetime dimensions.} It would also be interesting to study the OTOC in higher dimensional black holes. The gravitational modes responsible for scrambling in these black holes have been identified in \cite{Banerjee:2019vff}. It would be interesting to determine whether there is a similar saw-tooth behaviour in the OTOC and whether the overall growth of the OTOC is consistent with slow scrambling or whether the mode with the largest rate of growth comes to  dominate in that case.

\section*{Acknowledgments}
Some of the ideas on which this paper is based were discussed at the ``Microstates and Chaos'' panel of the Black-Hole Microstructure Conference (Saclay, 2020). BC thanks I.~Bena, E.~Martinec, D.~Stanford and N.~Warner for the organization and for discussions. BC also thanks S.~Khetrapal for collaboration on a related project.

This work is supported in part by FWO-Vlaanderen through project G006918N and by Vrije Universiteit Brussel through the Strategic Research Program ``High-Energy Physics.'' MDC is supported by a PhD fellowship from the Research Foundation Flanders (FWO). PH is supported by a PhD fellowship from the VUB Research Council. KN is supported by a Fellowship of the Belgian American Educational Foundation and by a grant from the John Templeton Foundation. CR is supported by a postdoctoral fellowship from the Research Foundation Flanders (FWO).
\appendix
\section*{Appendix}

\section{Oscillations and decay in the OTOC}
\label{app:oscillations}
\label{sec:OTOC_oscillations}
In this appendix, we describe in more detail the existence of an intermediate region of damped rapid oscillatory behaviour for the OTOC in non-rotating BTZ, which lies between the Lyapunov growth and the quasi-normal mode decay.
These damped oscillations appear before the quasi-normal mode decay whenever there is a hierarchy, controlled by the weight $h_V$, between the scrambling time $t_s$ and the time scale $t_{QN}$ at which the quasi-normal mode decay kicks in.

For a non-rotating BTZ black hole, the OTOC is given by ($h_W \gg h_V \gg 1$) \cite{Shenker:2014cwa}
\begin{align}
    \frac{\langle V(-i\epsilon_V,0) W(t-i\epsilon_W,x) V(i\epsilon_V,0) W(t+i \epsilon_W,x) \rangle_\beta }{ \langle V(-i\epsilon_V,0)V(i \epsilon_V,0)\rangle_\beta \langle W(t-i\epsilon_W,x) W(t+i\epsilon_W,x) \rangle_\beta}
    = \left( 1 - \frac{6 \pi i  h_W e^{\frac{2\pi}{\beta}\left(t-|x|\right)} }{c\sin \left( \frac{2\pi}{\beta} \epsilon_V \right) \sin \left( \frac{2\pi}{\beta} \epsilon_W \right)}  \right)^{-2h_V}\, ,
    \label{eq: SS OTOC}
\end{align}
where $h_V$ and $h_W$ are the conformal weights of the operators.
It is well known that \eqref{eq: SS OTOC} describes both the early-time Lyapunov growth and the quasi-normal mode decay of the OTOC.
In the following, we will discuss the presence of an intermediate regime where the OTOC shows rapid damped oscillations, with a decay that is doubly-exponential.

We start by noting that for $x \ll 1$
\begin{align}
    \log (1+x) \simeq x - \frac12 x^2 + O(x^3)\,.
\end{align}
Therefore, in the  regime where $\frac{6 \pi  h_W e^{\frac{2\pi}{\beta}\left(t-|x|\right)}}{c\sin \left( \frac{2\pi}{\beta} \epsilon_V \right) \sin \left( \frac{2\pi}{\beta} \epsilon_W \right)} \ll 1$, we can write \eqref{eq: SS OTOC} as
\begin{multline}
    \left( 1 - \frac{6 \pi i h_W e^{\frac{2\pi}{\beta}\left(t-|x|\right)}}{c\sin \left( \frac{2\pi}{\beta} \epsilon_V \right) \sin \left( \frac{2\pi}{\beta} \epsilon_W \right)} \right)^{-2h_V} \simeq
    \exp \left( i \frac{12 \pi  h_W h_V}{c\sin \left( \frac{2\pi}{\beta} \epsilon_V \right) \sin \left( \frac{2\pi}{\beta} \epsilon_W \right)} e^{\frac{2\pi}{\beta}\left(t-|x|\right)}\right)  \\[1em]
  \times \exp \left( -   \frac{36 \pi^2  h_W^2 h_V}{c^2\sin^2 \left( \frac{2\pi}{\beta} \epsilon_V \right) \sin^2 \left( \frac{2\pi}{\beta} \epsilon_W \right)} e^{\frac{4\pi}{\beta}\left(t-|x|\right)}
    +O\left( \frac{h_W^3 h_V e^{3t}}{c^3}\right) \right) \,.
    \label{eq: expansion otoc}
\end{multline}
Normally the second term in the exponent would be subleading, but since the first term is purely imaginary it does not contribute to the magnitude of the OTOC. Taking the real part of this expression, we obtain
\begin{equation}
    \text{Re} ({\rm OTOC}) \simeq \cos \left(\frac{12 \pi h_W h_V e^{\frac{2\pi}{\beta}\left(t-|x|\right)}}{c\sin \left( \frac{2\pi}{\beta} \epsilon_V   \right) \sin \left( \frac{2\pi}{\beta} \epsilon_W \right)}  \right) \exp \left( -   \frac{36 \pi^2  h_W^2 h_V}{c^2\sin^2 \left( \frac{2\pi}{\beta} \epsilon_V \right) \sin^2 \left( \frac{2\pi}{\beta} \epsilon_W \right)} e^{\frac{4\pi}{\beta}\left(t-|x|\right)}  \right)  \,.
\end{equation}
This starts very near 1 and decreases exponentially, as is well known. The first zero of the cosine, which corresponds to the time scale at which the commutator squared first becomes order 1 and is known as the scrambling time, $t_s$, occurs at
\begin{align}
    t_s \simeq |x| - \frac{\beta}{2\pi}\log \left( \frac{24 h_W h_V}{c\sin \left( \frac{2\pi}{\beta} \epsilon_V \right) \sin \left( \frac{2\pi}{\beta} \epsilon_W \right)} \right)\,.
\end{align}
After this time scale, the OTOC oscillates with a rapidly increasing frequency. At a time scale $t_d$ the OTOC starts to decay double-exponentially,
\begin{align}
    t_d \simeq |x| -  \frac{\beta}{2\pi} \log \left( \frac{6 \pi  h_W \sqrt{h_V}}{ c \sin \left( \frac{2\pi}{\beta} \epsilon_V \right) \sin \left( \frac{2\pi}{\beta} \epsilon_W \right)} \right)\,.
\end{align}
The separation between these timescales is controlled by the conformal weight of the lighter $V$ operator:
\begin{align}
    t_d - t_s \sim \frac{\beta}{\pi} \log h_V \,.
\end{align}
Finally at late times, $t \sim t_{QN}$, quasi-normal decay takes over
\begin{align}
    t_{QN} = |x| -  \frac{\beta}{2\pi} \log \left( \frac{6 \pi   h_W}{c \sin \left( \frac{2\pi}{\beta} \epsilon_V \right) \sin \left( \frac{2\pi}{\beta} \epsilon_W \right) } \right)\,.
\end{align}
In this regime, the OTOC decays exponentially at a rate controlled by $h_V/\beta$
\begin{align}
    {\rm OTOC} \simeq \left(\frac{6 \pi i  h_W}{ c \sin \left( \frac{2\pi}{\beta} \epsilon_V \right) \sin \left( \frac{2\pi}{\beta} \epsilon_W \right)} \right) ^{-2h_V} e^{-\frac{4 \pi h_V}{\beta}(t-|x|)}\,.
\end{align}
Note however that by this time, the OTOC has already been decaying doubly exponentially since $t_d$.
This time scale is separated from the earlier decay by the same amount as the scrambling time was separated from the doubly-exponential decay,
\begin{align}
    t_{QN} - t_d \sim  \frac{\beta}{\pi} \log h_V \,.
\end{align}

Although the geodesic approximation to the computation of the OTOC is not sensitive to the quasi-normal decay regime, we will show that in addition to the leading early-time behaviour given by
\begin{align}
    {\rm OTOC} \simeq e^{i\delta}\Big|_{\text{saddle}}\,,
    \label{eq: otoc app}
\end{align}
it correctly reproduces the first correction, i.e., the second factor in \eqref{eq: expansion otoc}.
The leading term is obtained by ignoring the back reaction of the geodesics on their trajectories and computing the dominating saddle point in \eqref{eq: otoc app} by a choice of point-splitting regulators together with condition \eqref{initial.velocity}, which is a local condition that does not include the effect of the eikonal phase factor. In \cite{Shenker:2014cwa}, the late-time quasi-normal type decay arises by taking the eikonal phase into account when computing the saddle point, which can be interpreted in terms of the propagation of the light particle ($V$) in the background of the heavier particle ($W$).
In the WKB approach used in this paper, a first correction to the OTOC in non-rotating BTZ can similarly be found by including the eikonal phase factor in the computing of the saddle of the $V$ particle, such that one finds\footnote{In \cite{Balasubramanian:2019stt}, a detailed derivation of the saddles was given (without including the eikonal phase factor). \eqref{eq: otoc expansion WKB} can be derived following those steps while including the effect of the eikonal phase.}
\begin{align}
     \frac{\langle V(0,0) W(t,x) V(-\epsilon_V,0) W(t- \epsilon_W,x) \rangle_\beta }{ \langle V(0,0)V( -\epsilon_V,0)\rangle_\beta \langle W(t,x) W(t-\epsilon_W,x) \rangle_\beta} \simeq
    &e^{ i \frac{\beta^2 G_N m_W m_V}{ 2\pi\varepsilon^2} e^{\frac{2\pi}{\beta}\left(t-|x|\right)}
    +   i \frac{\beta^4 G_N^2 m_W^2 m_V}{2 (2\pi)^3\varepsilon^3 \epsilon_V} e^{\frac{4\pi}{\beta}\left(t-|x|\right)}}  \, ,
    \label{eq: otoc expansion WKB}
\end{align}
with $m_i = 2 h_i$ and where $G_N$ is related to the central charge $c$ of the CFT by the Brown-Henneaux relation $c = \frac{3}{2G_N}$, in units where the AdS length is set to 1.
A few differences are to be noted with respect to \eqref{eq: expansion otoc}. First, a feature of our approach that is already present in the leading term, is that the (holographic) cutoff surface regulator $\varepsilon$ appears in the denominator of the eikonal phase instead of the time regulators. Second, the regulators $\epsilon_i$ are taken to be real shifts to the times (see \eqref{UV-regulator}). This type of time regulator leads to a correction to the phase of the OTOC and does not change its amplitude. In order to connect more directly with the result from \cite{Shenker:2014cwa}, we need to consider Euclidean time regulators\footnote{This is a natural choice from the CFT point of view, because the ordering of the operators inside the four-point function is fixed by the euclidean time ordering of the operators.} and send $\epsilon_i \rightarrow -i \epsilon_i$. Putting both types of regulators on the same footing, equations \eqref{eq: expansion otoc} and \eqref{eq: otoc expansion WKB} are in agreement, provided we consider small shifts $\epsilon_i$. As a consequence, the exact behaviour of the intermediate regime of the OTOC seems to be strongly dependent on the choice of regulator.

Note that the corrections to the leading early-time behaviour of the OTOC break the symmetry between $V$ and $W$, because a hierarchy was chosen between the two particles ($h_W \gg h_V$). As a result, the two series \eqref{eq: expansion otoc} and \eqref{eq: otoc expansion WKB} contain an increasing number of factors of $h_W$ together with a single factor of $h_V$.

\paragraph{Slow scrambling in the vacuum}
The OTOC has been computed for the vacuum state of a CFT$_2$ on an infinite line, in the vacuum block approximation \cite{Roberts:2014ifa}:
\begin{align}
    \frac{\langle V(-i\epsilon_V,0) W(t-i\epsilon_W,x) V(i\epsilon_V,0) W(t+i \epsilon_W,x) \rangle_\infty }{ \langle V(-i\epsilon_V,0)V(i \epsilon_V,0)\rangle_\beta \langle W(t-i\epsilon_W,x) W(t+i\epsilon_W,x) \rangle_\infty}
    = \left(1- \frac{6 \pi i h_W }{\epsilon_W \epsilon_V c} (t-|x|)^2\right)^{-2h_V}\,.
\end{align}

Again the same approximation scheme can be used in the regime where $\frac{6 \pi h_w }{\epsilon_W \epsilon_V c} (t-|x|)^2 \ll 1$ to write this as
\begin{align}
 \left(1- \frac{6 \pi i h_W }{\epsilon_W \epsilon_V c} (t-|x|)^2\right)^{-2h_V} \simeq
 e^{i \frac{12 \pi h_W h_V}{\epsilon_W \epsilon_V c} (t-|x|)^2 - \frac{36 \pi^2 h_W^2 h_V}{ \epsilon_W^2 \epsilon_V^2 c^2} (t-|x|)^4     + O \left(\frac{ h_W^3 h_V t^6}{c^3}\right)} \,.
\end{align}
Taking the real part of the OTOC, we see the same sort of damped oscillations, but with a power law instead of exponential dependence on $t$ in the argument,
\begin{align}
    {\rm Re}({\rm OTOC}) \simeq \cos\left(\frac{12 \pi h_W h_V}{\epsilon_W \epsilon_V c}  (t-|x|)^2 \right)
    e^{- \frac{36 \pi^2 h_W^2 h_V}{\epsilon_W^2 \epsilon_V^2 c^2} (t-|x|)^4}     \,.
\end{align}
The same time scales can be identified,
\begin{align}
    t_s \simeq |x| + \sqrt{\frac{ \epsilon_W \epsilon_V c}{24  h_W h_V} } \,, \\
    t_d \simeq |x| + \sqrt{\frac{  \epsilon_W \epsilon_V c}{6 \pi h_W \sqrt{h_V} } } \,,\\
    t_{QN} \simeq |x| + \sqrt{ \frac{\epsilon_W \epsilon_V c}{6 \pi h_W}  }\,,
\end{align}
where, once again, the weight $h_V$ controls the time separation between the different regimes.

\section{Center-of-mass energy in generic BTZ}
\label{app:center-of-mass}

In this appendix, we show that the center-of-mass energy of two colliding geodesics in a non-extremal BTZ black hole already contains an initial period of power law growth, followed by an exponential growth set by the temperature.
As the angular momentum of the black hole is tuned towards extremality, this region of power law growth extends to larger separation times until the exponential behaviour disappears completely at extremality.

The metric of a rotating BTZ black hole is given by
\begin{gather} \label{eqn:ds2extremalBTZ}
ds^2 = -f(r) dt^2 +\frac{dr^2}{f(r)} +r^2 \left( d\varphi - \frac{r_+ r_-}{ r^2} dt \right)^2\,, \quad f(r) = \frac{(r^2-r_+^2)(r^2-r_-^2)}{ r^2}\,,
\end{gather}
where the angular coordinate $\varphi$ is periodically identified, $\varphi \sim \varphi + 2\pi $. The angular velocity of the black hole is given by $\Omega = r_- / r_+$ and extremal BTZ corresponds to setting $r_+ = r_-$.

As explained in the main text, we consider a collision between radially infalling null geodesics.
One geodesic falls in at time $t_1$ and $\varphi=0$ and the other moves outwards and reaches the boundary at $t_2$ and the same $\varphi=0$.
The two cross at a radius $r_*$ and we will compute their center-of-mass energy at this point
\begin{align}
    s&=-2 m_{V} m_{W}\, k_{V} \cdot k_{W}\big|_{r=r_*},
\end{align}
where $k_{V}^\mu$ and $k_{W}^\mu$ are the velocities associated to the two geodesics.

This geometry has two Killing vectors, $\partial_t$ and $\partial_\varphi$, with associated conserved quantities
\begin{align}
E &= - \left(\partial_t \right)^\mu k_\mu
= \left( f(r) - \frac{r_+^2 r_-^2}{ r^2} \right) \dot{t} +  r_+ r_- \dot{\varphi}\,,\\
L &= \left(\partial_\varphi \right)^\mu k_\mu
= r^2 \dot{\varphi} - r_+ r_- \dot{t}\,.
\end{align}

Setting $L=0$ and using the null condition $k^2=0$, the final coordinate can be fixed to be
\begin{equation}
\dot{r}^2  =  E^2,
\end{equation}
so that the center-of-mass energy of two geodesics with $L_i=0$ and $E_V=E_W=E$, at the collision, can be written in terms of only the energy and the position of the collision,
\begin{gather}
s = \frac{4m_V m_W  E^2 }{  f(r_*)}.
\end{gather}
The horizon is located at the outermost zero of $f(r)$.
For non-extremal BTZ, this is a simple zero, whereas for extremal BTZ it becomes a double zero. For $r_* \sim r_+$, the center-of-mass energy can be approximated by
\begin{align}
    s &\sim \frac{2 m_V m_W r_+ E^2}{(r_+^2 - r_-^2) (r_*-r_+)} \,,
     &r_*^2 - r_+^2 \ll r_+^2 - r_-^2\,,\\
    &\sim \frac{m_V m_W E^2}{(r_*-r_+)^2} \,,  &r_+^2 - r_-^2 \ll r_*^2 - r_+^2 \ll r_+^2\,.
\end{align}

The remaining task is to relate the position of the collision to the location at which the geodesics meet the boundary,
\begin{gather}
 t_{\infty} - t_* = \int^{\infty}_{r_*} \frac{dt}{dr} dr =  \int^{\infty}_{r_*} \frac{\dot{t}}{\dot{r}} dr\, = \pm \int^{\infty}_{r_*}  \frac{dr}{f(r)} \,.
\end{gather}
Once again, we distinguish two regimes, depending on the value of the ratio of the distance between the two horizons and the distance between the collision and the outer horizon,
\begin{align}
\int_{r^*}^\infty \frac{dr}{f(r)}
&\sim \frac{r_+}{2 ( r_+^2 - r_-^2)} \int_{r_*}  \frac{dr}{r-r_+} =  -\frac{r_+ \log(r_*-r_+)}{2 ( r_+^2 - r_-^2)} \,, \quad &r_*^2 - r_+^2 \ll r_+^2 - r_-^2\,,
\label{eqn:ext_vs_nonext_BTZ_1} \\
&\sim \frac{1}{4} \int_{r_*}  \frac{dr}{(r-r_+)^2} = \frac{1}{4  (r_*-r_+)}\,, \quad &r_+^2 - r_-^2 \ll r_*^2 - r_+^2\ll r_+^2\,.
\label{eqn:ext_vs_nonext_BTZ_2}
\end{align}
In other words, for non-extremal black holes we  find that the depth inside the geometry explored by probes separated by a distance $\Delta t$ on the boundary scales like \begin{align}
r_* -r_+ &\sim e^{-\frac{ r_+^2 - r_-^2}{r_+} \Delta t} \,,  &r_*^2 - r_+^2 \ll r_+^2 - r_-^2  \,,\\
&\sim  \frac{1}{2 \Delta t} \,, &
r^2_+ - r_-^2 \ll r_*^2-r_+^2 \ll r_+^2 \,.
\label{eqn:t_of_r*}
\end{align}
This leads to
\begin{align}
    s &\sim \frac{2 m_V m_W  r_+ E^2 }{(r_+^2 - r_-^2)}
    e^{\frac{ r_+^2 - r_-^2}{r_+} \Delta t}\,,
     &r_*^2 - r_+^2 \ll r_+^2 - r_-^2\,,\\
    &\sim 4 m_V m_W  E^2  \Delta t^2 \,,  &r_+^2 - r_-^2 \ll r_*^2 - r_+^2 \ll r_+^2 \,.
\end{align}
Note that for a non-rotating black hole, $r_-=0$, and the second regime never applies. In that case, $r_*$ approaches the horizon exponentially.
As the spin of the black hole increases, a region of power law approach to the horizon appears before the exponential approach begins. As the extremal limit is approached this power law regime persists for longer until at extremality, $r_-=r_+$, the exponential regime disappears completely and only the power law regime remains.

\section{Shock wave in extremal BTZ}\label{appendix:shock wave}
Following the method presented in \cite{Balasubramanian:2019stt}, we construct the shock wave solutions used in the main text, which are associated to geodesics with zero angular momentum in extremal BTZ. We make use of the fact that extremal BTZ is just a patch of pure $AdS_3$, where the latter may be viewed as a hyperboloid in four-dimensional flat space through the constraint
\begin{equation}
\label{hyperboloid}
\eta_{MN} X^M X^N=-1, \qquad \eta_{MN}=\text{diag}\left(-1,1,1,-1\right).
\end{equation}
We refer to $X^M$ as `embedding coordinates'. We then choose the following set of independent lightcone coordinates,
\begin{equation}
\mathcal{V}=X^0+X^1, \qquad \mathcal{U}=X^0-X^1, \qquad \mathcal{Z}=X^2,
\end{equation}
while the remaining embedding coordinate $X^3$ is determined from either one of the two branches
\begin{equation}
X^3=\pm \sqrt{1-\mathcal{U}\mathcal{V}+\mathcal{Z}^2}.
\end{equation}
A choice of branch corresponds to a choice of either one of the two `hemispheres' of the AdS hyperboloid.

A highly energetic particle following a null trajectory along $\mathcal{V}=\mathcal{Z}=0$ in the lower hemisphere $(X^3=-1)$ has a stress tensor given by
\begin{align}
T_{\mathcal{V}\mathcal{V}}=-m k_{\mathcal{V}}\, \delta(\mathcal{V}) \delta(\mathcal{Z})\Theta(-X^3),
\end{align}
where its velocity $k_\mathcal{V}$ is a constant of motion. The Heaviside step function $\Theta(-X^3)$ explicitly restricts the source to lie in the lower hemisphere of the AdS hyperboloid. It has the effect of discarding a second null geodesic at $\mathcal{V}=\mathcal{Z}=0$, but lying in the upper hemisphere $(X^3=1)$ instead. When going to the extremal BTZ patch, this second geodesic simply coincides with the `reflected' continuation at the AdS conformal boundary of the geodesic of interest. This will be made explicit later on. Here, we are interested in null geodesics either created or absorbed at the AdS conformal boundary such that we don't consider such boundary `reflections'. The associated shock wave geometry is found by solving Einstein's equations sourced by the above stress tensor. It takes the form
\begin{equation}
\label{shock wave}
ds^2=ds^2_{AdS}+ds^2_{SW}, \qquad ds^2_{SW}=-16\pi G_N m k_{\mathcal{V}}\ \Pi(\mathcal{Z})\delta(\mathcal{V})\Theta(-X^3)d\mathcal{V}^2,
\end{equation}
where $\Pi(\mathcal{Z})$ solves
\begin{equation}
\left[\left(1+\mathcal{Z}^2\right)\partial_\mathcal{Z}^2+\mathcal{Z} \partial_\mathcal{Z}-1\right]\Pi(\mathcal{Z})=-\delta(\mathcal{Z}).
\label{eq: diff eq Pi}
\end{equation}
Before solving this equation, we first need to determine the appropriate boundary conditions.

The shock wave sourced by a highly energetic particle in any locally $AdS_3$ spacetime can be mapped to the solution \eqref{shock wave} written in embedding coordinates $X^M$. In particular, shock waves in maximally rotating BTZ black hole backgrounds may be found by appropriate coordinate transformation. To achieve this, we first give the map\footnote{It may be found by performing the following chain of coordinate transformations: $\bar{X}^M \to$ Poincar\'e patch of AdS $\to (t,r,\varphi) \to (v,r,\phi)$; see \cite{Gralla:2019isj}.} between some other \textit{intermediate} embedding coordinate system $\bar{X}^M$
and retarded BTZ coordinates $(v,r,\phi)$,
\begin{subequations}
\label{Xbar-coord}
\begin{align}
    \bar{X}^0 &= \frac{1}{2} e^{-r_+ \phi } (1+(r+r_+) (2 v+\phi ))-\frac{(r-r_+) e^{r_+ \phi }}{4 r_+}, \\
    \bar{X}^1 &=\frac{1}{2} e^{-r_+ \phi }
   (1+(r+r_+) (2 v+\phi ))+\frac{(r-r_+) e^{r_+ \phi }}{4 r_+}, \\
   \bar{X}^2 &=\frac{e^{r_+ \phi } (1+(r-r_+) (2 v+\phi ))}{4 r_+}-\frac{1}{2}
   (r+r_+) e^{-r_+ \phi },\\
    \bar{X}^3 &= \frac{e^{r_+ \phi } (1+(r-r_+) (2 v+\phi ))}{4r_+}+\frac{1}{2} (r+r_+) e^{-r_+ \phi }.
\end{align}
\end{subequations}
Any null geodesic in extremal BTZ then maps to a null ray in the intermediate coordinate system $\bar{X}^M$. Depending on the null geodesic of interest, as a final step we need to find the isometry relating these intermediate coordinates $\bar{X}^M$ to the coordinates $X^M$ described above, in such a way that the null geodesic lies at $\mathcal{V}=\mathcal{Z}=0$ in the lower hemisphere of the AdS hyperboloid $(X^3=-1)$.

\paragraph{Outgoing shock wave.} The trajectory of an outgoing geodesic with zero angular momentum has been described in section~\ref{section:geodesics}. It can be mapped to the null ray
\begin{equation}
\label{geo-embedding}
\mathcal{V}=\mathcal{Z}=0, \qquad (X^3=-1),
\end{equation}
through the following AdS isometry,
\begin{equation}
\label{isometry}
\begin{pmatrix}
X^0\\
X^1\\
X^2\\
X^3
\end{pmatrix}
=
\begin{pmatrix}
1 & 0 & 0 & 0\\
0 & 1 & 0 & 0\\
0 & 0 & \cosh c_2 & \sinh c_2\\
0 & 0 & \sinh c_2 & \cosh c_2
\end{pmatrix}
\begin{pmatrix}
a & 0 & 0 & -b\\
0 & 1  & 0 & 0\\
0 & 0 & 1 & 0\\
b & 0 & 0 & a\\
\end{pmatrix}
\begin{pmatrix}
1 & 0 & 0 & 0\\
0 & 1 & 0 & 0\\
0 & 0 & \cosh c_1 & \sinh c_1\\
0 & 0 & \sinh c_1 & \cosh c_1
\end{pmatrix}
\begin{pmatrix}
\bar{X}^0\\
\bar{X}^1\\
\bar{X}^2\\
\bar{X}^3
\end{pmatrix},
\end{equation}
with parameters
\begin{subequations}
\begin{align}
a&=\frac{e^{2r_+ \phi_{out}}-2r_+(2v_{out}+\phi_{out})}{e^{2r_+ \phi_{out}}+2r_+(2v_{out}+\phi_{out})},\qquad b=\frac{e^{r_+ \phi_{out}}\sqrt{8r_+(2v_{out}+\phi_{out})}}{e^{2r_+ \phi_{out}}+2r_+(2v_{out}+\phi_{out})},\\
c_1&=- r_+ \phi_{out} -\frac{1}{2} \log \frac{2v_{out}+\phi_{out}}{2 r_+},\qquad c_2=\frac{1}{2} \log 2r_+(2v_{out}+\phi_{out}).
\end{align}
\end{subequations}
One may check that $a^2+b^2=1$ such that the associated isometry is a rotation in the $X^0$-$X^3$ plane. Using \eqref{Xbar-coord} and \eqref{isometry}, we get a final expression for the null coordinate $\mathcal{V}$,
\begin{equation}
\label{coord-transf V}
\mathcal{V} = \frac{e^{-r_+ \Delta \phi}\left[1+(r-r_+)(2\Delta v+\Delta \phi)\right]-e^{r_+ \Delta \phi}\left[1+(r+r_+)(2\Delta v+\Delta \phi)\right]}{e^{r_+ \phi_{out}}+2r_+(2v_{out}+\phi_{out})e^{-r_+ \phi_{out}}},
\end{equation}
with $\Delta \phi\equiv \phi_{out}-\phi$ and $\Delta v \equiv v_{out}-v$. We can now explicitly check that the null ray lying at $\mathcal{V}=\mathcal{Z}=0$ in the upper hemisphere $(X^3=1)$ maps to a `reflected' ingoing null geodesic with zero angular momentum and trajectory $v(r)=v_{out},\ \phi(r)=\phi_{out}$.

\begin{figure}[h!]
\centering
\subfloat[]{{\includegraphics[scale=0.55]{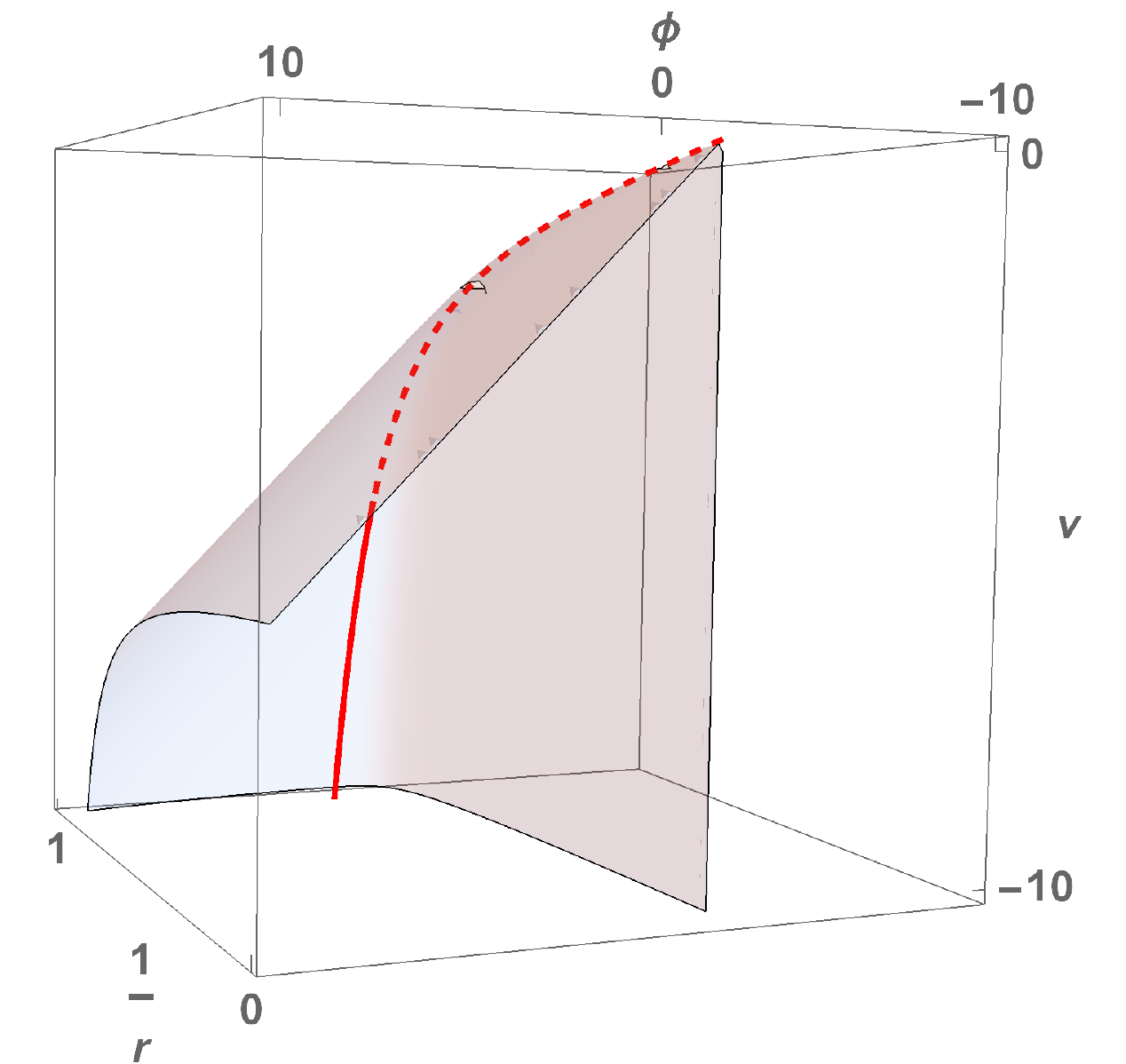} }}\hskip 5mm
\subfloat[]{{\includegraphics[scale=0.55]{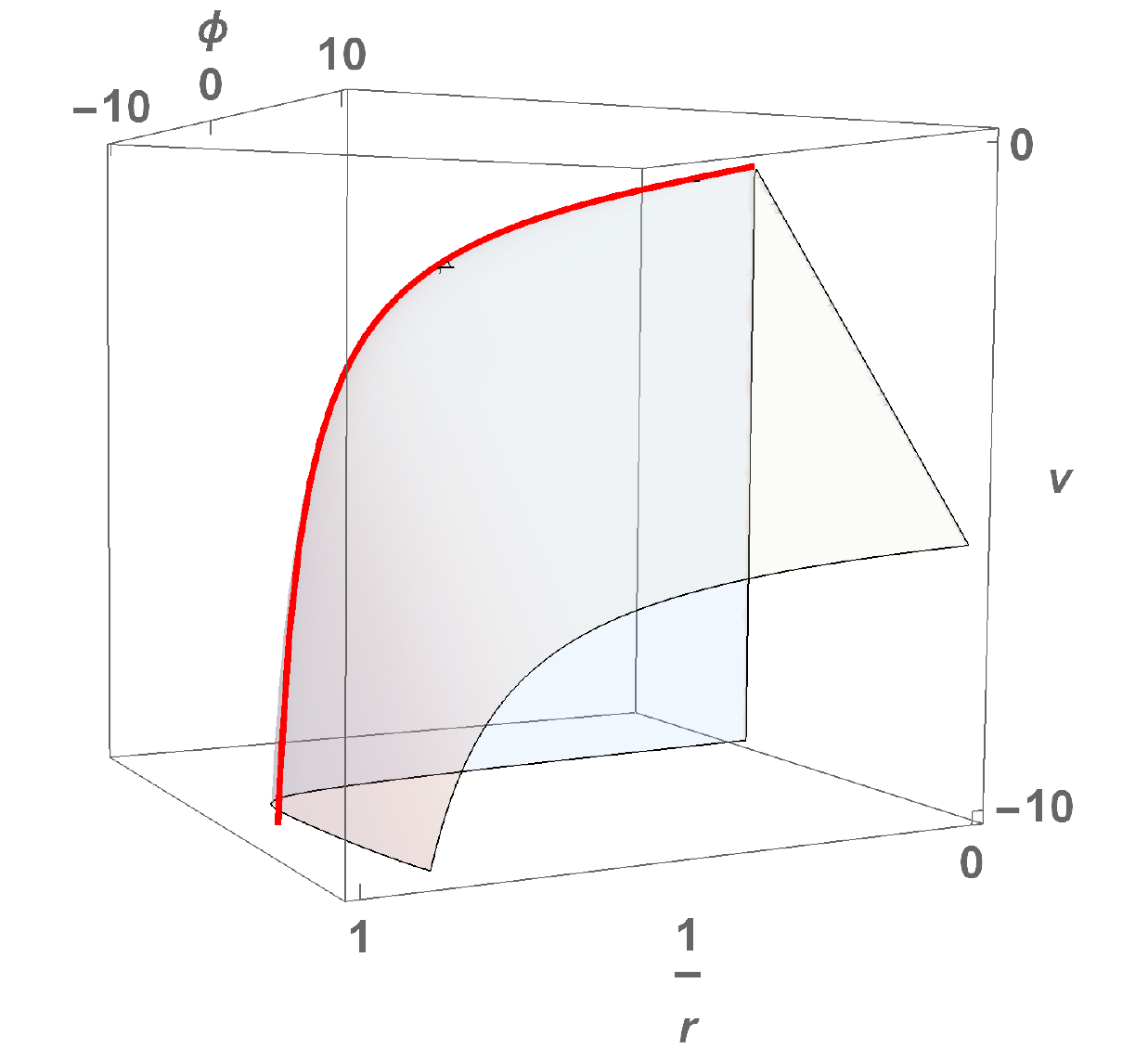} }}\caption{Support of the shock wave \eqref{shock wave-outgoing} emitted by a null outgoing geodesic with boundary insertion point $v_{out}=0,\,  \phi_{out}=-2$, viewed in retarded coordinates $(v,r,\phi)$ from two different perspectives. The horizon radius has been set to $r_+=1$ and the outgoing geodesic is indicated by the red line. The shock wave is emitted from the geodesic towards the conformal boundary. The shock wave \eqref{eq: shock wave-outgoing compactified} associated with a conformal boundary with cylinder topology used in the main text, is found by adding images in order to implement the periodic identification $\phi \sim \phi + 2\pi$.}
\label{fig: shock wave 3D}
\end{figure}

In the main text, we only need the $rr$-component of the shock wave created by the outgoing geodesic. It is found from \eqref{shock wave} by use of the coordinate transformation \eqref{coord-transf V}, and by plugging the relation between one component of the velocity \eqref{eq: momentum W} and the conserved energy $E$ defined in \eqref{eq:conserved quantities},
\begin{equation}
k_r=\frac{2r^2}{\left(r^2-r_+^2\right)^2}\, E,
\end{equation}
yielding
\begin{align}
\label{shock wave-outgoing}
h_{rr}&=\frac{8 \pi G_N m E}{r_+}\, (2\Delta v+\Delta \phi)\sinh (r_+\Delta \phi) \Pi(\mathcal{Z}) \delta \left(r- r_0(v,\phi) \right) \Theta(-X^3),
\end{align}
with
\begin{align}
\label{r0}
r_0(v,\phi) &=\frac{1}{2\Delta v+ \Delta \phi}-r_+ \coth r_+\Delta \phi,
\end{align}
and
\begin{subequations}
\begin{align}
\mathcal{Z}(v,r_0(v,\phi),\phi)&=- \frac{\sinh^2(r_+ \Delta \phi) -  r_+^2 \left(2 \Delta v +\Delta \phi \right)^2}{2  r_+ \sinh( r_+ \Delta \phi) \left(2 \Delta v +\Delta \phi \right) },\\
X^3\left(v,r_0(v,\phi),\phi\right)&=\frac{\sinh^2 (r_+\Delta \phi)+r_+^2\left(2\Delta v+\Delta \phi\right)^2}{2r_+\sinh(r_+\Delta\phi)\left(2\Delta v+\Delta \phi\right)}.
\end{align}
\end{subequations}
The restriction $X^3<0$ in \eqref{shock wave-outgoing} is equivalent to $\Delta \phi (2\Delta v+\Delta \phi)<0$. From \eqref{r0}, we find that only $-2\Delta v<\Delta \phi<0$ yields a positive value for $r_0$. Since the BTZ radial coordinate $r$ ranges over positive values, only a positive $r_0$ can contribute to the support of the shock wave \eqref{shock wave-outgoing}. The support of this outgoing shock wave is depicted in figure~\ref{fig: shock wave 3D}.

Let us now come back to the boundary conditions that one needs to impose on \eqref{eq: diff eq Pi}. The support of the shock wave stretches out from the outgoing geodesic towards the boundary and reaches the boundary when $ \Delta \phi  = 0$ or $\Delta \phi = -2 \Delta v$, as may be seen from \eqref{r0}. Those values correspond to $\mathcal{Z} \rightarrow \pm \infty $ in embedding space. We therefore impose that $\Pi(\mathcal{Z})$ should decay at infinity. Together with a continuity constraint on $\Pi(\mathcal{Z})$ at $\mathcal{Z} = 0$, one finds the following solution
\begin{equation}
\Pi(\mathcal{Z}(v,r_0(v,\phi),\phi)) =
  \begin{cases}
       \frac{1}{2} \left( \sqrt{1+\mathcal{Z}^2}+\mathcal{Z} \right) =- \frac{\sinh r_+ \Delta \phi }{2r_+( 2\Delta v+ \Delta \phi)} & \text{for } \mathcal{Z}\leq 0 \\
      \frac{1}{2} \left( \sqrt{1+\mathcal{Z}^2}-\mathcal{Z} \right) =- \frac{r_+(2\Delta v+ \Delta \phi)}{2\sinh r_+\Delta \phi} & \text{for } \mathcal{Z}>0,
  \end{cases}
\end{equation}
where we assumed $-2\Delta v < \Delta \phi < 0$ in the last equality.

Equation \eqref{shock wave-outgoing} represents the $rr$-component of the shock wave of an outgoing null geodesic in the decompactified limit. When the coordinate $\phi$ is periodic with period $2\pi$, one should sum over images,
\begin{align}
\label{eq: shock wave-outgoing compactified}
h_{rr}^{(2\pi)} &= \sum_{n\in \mathbb{Z}} h_{rr}(r,\Delta v,\Delta \phi_n), \qquad \Delta \phi_n \equiv \phi_{out} - \phi + 2\pi n.
\end{align}
The shock wave of a single image of the geodesic only has support at bulk points $(v,r_0(v,\phi),\phi)$ satisfying $-2\Delta v < \Delta \phi < 0$, such that only finitely many images will contribute to the shock wave considered at any given bulk point.

\paragraph{Ingoing shock wave.} The derivation of the shock wave sourced by an ingoing particle with zero angular momentum is identical, provided that one uses advanced coordinates \eqref{advanced-coord}. The relevant $rr$-component is thus obtained from \eqref{eq: shock wave-outgoing compactified} by making the replacements
\begin{equation}
\Delta v\, \to\, \Delta u\equiv u-u_{in}, \qquad \Delta \phi_n\, \to\, \Delta \phi'_n\equiv \phi'-\phi'_{in}+2\pi n.
\end{equation}

\paragraph{Comparison with earlier literature.} It is worth contrasting the shock wave expression \eqref{eq: shock wave-outgoing compactified} to the one obtained in the case of non-extremal BTZ in \cite{Mezei:2019dfv}; see also \cite{Jahnke:2019gxr}. These authors considered the shock waves sourced by geodesics lying on the past and future horizons of a non-maximally rotating BTZ black hole, respectively. As already mentioned, this allows one to compute the leading contribution to the OTOC in the regime $\Delta t\gg \beta$ but precludes the study of zero temperature states which include vacuum AdS and extremal BTZ. For these, it is crucial that one considers shock waves sourced by geodesics away from horizons instead. In particular, \eqref{eq: shock wave-outgoing compactified} cannot be straightforwardly obtained as the zero-temperature limit of the shock wave presented in \cite{Jahnke:2019gxr,Mezei:2019dfv},
which simply diverges. A similar qualitative feature, however, is the appearance of a sum over images due to the angular periodicity. Each image may be thought of as being associated to an additional winding of the shock wave around the black hole. More images are being generated as the time separation $\Delta v$ with the source particle increases. In particular, an infinite number of images contribute to \eqref{eq: shock wave-outgoing compactified} in the limit $\Delta v\to \infty$. The shock wave described in \cite{Mezei:2019dfv} may be similarly expressed as an infinite sum over images, although it has been explicitly resummed in that case. We refer the reader to section~\ref{subsection: eikonal phase} for further comments on the relation between shock waves in maximally and non-maximally rotating BTZ black holes.


\providecommand{\href}[2]{#2}\begingroup\raggedright\endgroup

\end{document}